\documentclass{article}

\usepackage{arxiv}

\usepackage[utf8]{inputenc} % allow utf-8 input
\usepackage[T1]{fontenc}    % use 8-bit T1 fonts
\usepackage{hyperref}       % hyperlinks
\usepackage{url}            % simple URL typesetting
\usepackage{booktabs}       % professional-quality tables
\usepackage{amsfonts}       % blackboard math symbols
\usepackage{nicefrac}       % compact symbols for 1/2, etc.
\usepackage{microtype}      % microtypography
\usepackage{lipsum}		% Can be removed after putting your text content
\usepackage{graphicx}
\usepackage{natbib}
\usepackage{doi}

\usepackage{subcaption}
\usepackage{amsmath}

\usepackage{textcomp, gensymb}

\title{Stochastic modeling of physical drag coefficient - its impact on orbit prediction and space traffic management}

\author{Smriti Nandan Paul \\
	Department of Mechanical and Aerospace Engineering\\
	West Virginia University\\
	Morgantown, WV 26505 \\
	\texttt{smritinandan.paul@mail.wvu.edu} \\
	\And
	{Phillip Logan Sheridan} \\
	Department of Mechanical and Aerospace Engineering\\
    West Virginia University\\
	Morgantown, WV 26505 \\
	\texttt{pls.sheridan@gmail.com} \\
	\And
	{Richard J. Licata} \\
	Department of Mechanical and Aerospace Engineering\\
    West Virginia University\\
	Morgantown, WV 26505 \\
	\texttt{rjlicata@mix.wvu.edu} \\	
	\And
	{Piyush M. Mehta} \\
	Department of Mechanical and Aerospace Engineering\\
    West Virginia University\\
	Morgantown, WV 26505 \\
	\texttt{piyush.mehta@mail.wvu.edu} \\		
}

\renewcommand{\shorttitle}{}

\begin{document}
\maketitle

\begin{abstract}
	Ambitious satellite constellation projects by commercial entities and the ease of access to space in recent times have led to a dramatic proliferation of low-Earth space traffic. It jeopardizes space safety and long-term sustainability, necessitating better space traffic management (STM). Correct modeling of uncertainties in force models and orbital states, among other things, is an essential part of STM. For objects in the low-Earth orbit (LEO) region, the uncertainty in the orbital dynamics mainly emanate from limited knowledge of the atmospheric drag-related parameters and variables. In this paper, which extends the work by \cite{PSMH2021}, we develop a feed-forward deep neural network model for the prediction of the satellite drag coefficient for the full range of satellite attitude (i.e., satellite pitch $\in$ ($-90\degree$, $+90\degree$) and satellite yaw $\in$ ($0\degree$, $+360\degree$)). The model simultaneously predicts the mean and the standard deviation and is well-calibrated. We use numerically simulated physical drag coefficient data for training our neural network. The numerical simulations are carried out using the test particle Monte Carlo method using the \textit{diffuse reflection with incomplete accommodation} gas-surface interaction model. Modeling is carried out for the well-known CHAllenging Minisatellite Payload (CHAMP) satellite. Finally, we use the Monte Carlo approach to propagate CHAMP over a three-day period under various modeling scenarios to investigate the distribution of radial, in-track, and cross-track orbital errors caused by drag coefficient uncertainty. 
\end{abstract}

% keywords can be removed

\section{Introduction}
\label{sec1}
Since the launch of the first artificial satellite, Sputnik, and up until recently, the near-Earth space environment saw a nearly balanced, steady growth of objects. However, recent advancements in commercially viable space technologies, satellite mega-constellation launches from private aerospace companies, and access to launch vehicles offering secondary payload services have led to unsustainable population growth. \cite{lemmens2020space} investigate conjunction events for the European Space Agency (ESA) satellites at low altitude LEO. As highlighted in their study, an increasing and significant portion of those close encounters are due to satellite constellations and small satellites. The LEO population growth, likely to go unabated, warrants better modeling of dynamical uncertainties and a more accurate prediction of orbital errors to make more informed decisions about space traffic management (STM) functions such as satellite conjunction occurrence and maneuvers for collision avoidance. The conservative perturbation forces are well-modeled, and the prime source of dynamical uncertainty for an LEO object is the atmospheric drag. Other non-conservative forces, such as solar radiation pressure (SRP), can also be a source of significant dynamical uncertainty for high area-to-mass ratio (HAMR) objects, but this study focuses only on ``typical" space objects which are low area-to-mass ratio (LAMR). 

For a satellite with mass $m$, the acceleration due to atmospheric drag is given by the following commonly accepted equation:
\begin{equation}
    \vec{a}_{D} = -\frac{1}{2}\rho \frac{C_D A_{proj}}{m}v_{rel}\vec{v}_{rel}
\end{equation}
where $\rho$ is the atmospheric density, $C_D$ is the drag coefficient, $A_{proj}$ is the projected area of the satellite perpendicular to the flow direction, $\vec{v}_{rel}$ is the velocity of the satellite relative to the atmosphere, and $v_{rel}$ is the magnitude of $\vec{v}_{rel}$. Except for the satellite mass, which is usually known from the operators and constant for a non-maneuvering satellite, all other parameters can have uncertainties. The uncertainty in the relative velocity $\vec{v}_{rel}$ arises from the uncertainty in local thermospheric winds, which can be as high as several hundreds of meters per second \citep{gaposchkin1988analysis}. Barring a perfectly spherical satellite, the uncertainty in the projected area emanates from missing/uncertain satellite attitude data and uncertainty in the computation of relative velocity vector direction. Much of the literature focuses on modeling the atmospheric density $\rho$, which has a complex dependency upon parameters such as geomagnetic indices, solar flux, the composition of the atmosphere, data measured by onboard satellite instrumentation, and others. These parameters themselves have uncertainties, which ultimately equate to uncertainty in the atmospheric density estimation. Instead of focusing on the prevalent density modeling, the current paper focuses on modeling the drag coefficient, a parameter that captures the interaction between the satellite surface and the atmospheric particles. Like density, modeling the drag coefficient is an involved task because of its dependency on atmospheric composition, satellite and atmospheric temperatures, and atmospheric winds.

Broadly speaking, drag coefficient modeling falls under one of the following three categories - fixed, fitted, and physical \citep{MEHTA2022}. When using the fixed approach, the drag coefficient is considered to be constant. In the fitted approach, the drag coefficient is estimated using a filtering method as part of an orbit determination process. In this study, our focus is on modeling the physical drag coefficient, which is determined by simulating the exchange of energy and momentum between the surface of the spacecraft and free-stream atmospheric particles \citep{chambre2017flow}. The most common practice for physical drag coefficient determination is to use computationally expensive numerical methods such as the Panel method, Ray-tracing Panel (RTP) method, Test Particle Monte Carlo (TPMC) method, or the Direct Simulation Monte Carlo (DSMC) method \citep{MOSTAZAPRIETO201456}. To avoid the high computational cost of the numerical methods, \cite{MEHTA20141590} build surrogate models based on Gaussian Process Regression (GPR) \citep{Rasmussen2004} for predicting satellite drag coefficient. The authors demonstrate that GPR is able to accurately predict drag coefficient with root mean square percentage errors below $1\%$ for a number of simple and complex geometries. However, \cite{MEHTA20141590} do not carry out uncertainty quantification. Building upon the work by \cite{MEHTA20141590}, \cite{PSMH2021} develop GPR and Monte Carlo Dropout \citep{pmlr-v48-gal16} based feed-forward deep neural network (FFDNN) models for stochastic/probabilistic prediction of the satellite drag coefficient. The authors demonstrate that both models are able to produce reasonably accurate and well-calibrated drag coefficient estimates. However, their models are valid for only a limited attitude range with satellite pitch and yaw varying between $-10\degree$ and $+10\degree$. The current paper aims to develop stochastic models for drag coefficient prediction for the full attitude range, i.e., satellite pitch $\in$ ($-90\degree$, $+90\degree$) and satellite yaw $\in$ ($0\degree$, $+360\degree$). Additionally, \cite{PSMH2021} use a mere 1000 ensemble points for training their machine learning models, which is not sufficient for the full attitude range. Tens of thousands of ensemble training points \citep{doi:10.1137/18M1170157} are desirable to accurately capture the full-attitude drag coefficient variations. Because GPR scales poorly with the data size, it is computationally infeasible to use GPR for the full-attitude drag coefficient modeling. Partially scalable variants of GPR are discussed later in the paper. Machine learning models such as the Monte Carlo Dropout-based FFDNN \citep{PSMH2021} or other variants of the neural network are scalable and can also provide an uncertainty estimate, making them ideal for large-scale full-attitude drag coefficient modeling.

In this paper, we use an in-house developed FFDNN model that directly predicts the mean and the standard deviation. We prefer the direct prediction of the uncertainty against an ensemble approach like the Monte Carlo dropout technique for uncertainty quantification because of the lesser computational costs in the direct prediction method. For any regression, the quality and quantity of the training data are critical to the correctness of the developed machine learning model. For machine learning training purposes in the current study, high-quality drag coefficient data are generated using the numerical TPMC method \citep{davis61}. The gas-surface interaction (GSI) in the TPMC simulations is modeled using the Diffuse Reflection With Incomplete Accommodation (DRIA) model \citep{Walker2014a, Walker2014b}. 

The current paper has a number of desirable goals: (1) the developed drag coefficient prediction model must be quick to evaluate and valid for any satellite orientation, (2) the developed model should be accurate not only in terms of the mean predictions but also provide meaningful and reliable uncertainty estimates, (3) the data size or the number of samples used to train the models should be sufficiently large to capture the drag coefficient variations with sufficient accuracy, and (4) the developed models should be inductive (as opposed to transductive) so that they can be saved and later re-used in an orbit propagation framework and perform orbital perturbation studies in a computationally efficient manner.

We organize the remainder of this paper into the following sections: section 2 provides brief background knowledge about the TPMC method and the DRIA GSI model. Section 3 provides details about the data used to train the regression models. We discuss the machine learning model used for the drag coefficient prediction in section 4. Model calibration, which is a concept closely related to the reliability of the predicted uncertainty, is discussed in section 5. Prediction performance for different data sizes is analyzed in section 6. Section 7 highlights difficulties in using scalable Gaussian processes for large-scale drag coefficient modeling. We investigate the effects of drag coefficient uncertainty on the orbital state uncertainties in section 8. Finally, we summarize the paper and provide the conclusions in the last section.

\section{Background - TPMC and the DRIA GSI Model}
The TPMC is a numerical method for computing the physical drag coefficient in the free molecular flow (FMF) regime. In TPMC, test particles representing actual molecules are sequentially fired into the computational domain. Each particle is fired with a probabilistically determined velocity. The TPMC assumes that molecules do not collide with one another, which speeds up computations while maintaining accuracy on par with other Monte Carlo techniques like the DSMC. The TPMC method is versatile because it can simulate different GSI models and easily handle flows with complex boundaries. 

As stated, our TPMC simulations are carried out using the DRIA GSI model. In the DRIA GSI model, the reflected particles have a diffuse angular distribution, based on Knudsen's cosine law \citep{knudsen1916}. The particles may exchange energy with the surface depending on the value of the energy accommodation coefficient. For more details on the DRIA GSI model, refer to \cite{ Walker2014a, Walker2014b, pilinski2010, moe2004simultaneous, sutton2009normalized}.

\section{Input Data for the Machine Learning Models}  
The training data for our predictive models are generated by numerical computation of the physical drag coefficient using the TPMC method, which is implemented using the West Virginia University (WVU) Response Surface Modeling (RSM) toolkit \citep{SHERIDAN20223828}. The WVU RSM toolkit is an open-source software package and can be accessed at \url{https://github.com/ASSISTLaboratory/WVU_RSM_Suite}. In this paper, drag coefficient modeling is carried out for the following primary LEO atmospheric constituents - atomic hydrogen ($H$), helium ($He$), atomic nitrogen ($N$), molecular nitrogen ($N_2$), atomic oxygen ($O$), and molecular oxygen ($O_2$). The total drag coefficient can be computed from the drag coefficients of the constituent species.

For the TPMC method with the DRIA GSI model, six independent variables determine the value of the dependent drag coefficient - (i) relative velocity of the satellite, $v_{\infty}$, (ii) satellite surface temperature, $T_w$, (iii) atmospheric translational temperature, $T_{\infty}$, (iv) energy accommodation coefficient, $\alpha$, (v) satellite yaw, $\beta$, and, (vi) satellite pitch, $\Phi$. In this study, the input configurations for the numerical simulations are carefully selected via the Latin Hypercube sampling (LHS) method \citep{MBC1979}. We use a high-fidelity geometry model corresponding to the CHAllenging Minisatellite Payload (CHAMP) satellite for our analysis. A total of 50,000 LHS design points are selected for each of the species $H$, $He$, $N$, $N_2$, $O$, $O_2$ for training purposes. The upper and lower bounds defining the LHS design points are given in Table \ref{LHS_bounds}.

\begin{table}[htbp]
    \centering 
    \caption{Upper and lower bounds defining the LHS design points.}
   \label{LHS_bounds}
   \begin{tabular}{|l | l | l |} % Column formatting, 
      \hline
      Independent Variables & Lower Bound & Upper Bound\\
      \hline 
      $v_{\infty}$      & 7250.0 m/s & 8000.0 m/s \\
      \hline
         $T_w$       & 100.0 K     &  2000.0 K \\
         \hline
      $T_{\infty}$       & 200.0 K  & 2000.0 K \\
      \hline
      $\alpha$       & 0.0  & 1.0 \\
      \hline
       $\beta$ & $0\degree$   &  $360\degree$ \\
       \hline
       $\Phi$ & $-90\degree$   &  $90\degree$ \\
      \hline
   \end{tabular}
\end{table}

The number of training samples, i.e., 50,000, is determined through a data size sensitivity analysis, which is discussed later. Besides the training data points, a different set of 50,000 LHS points, using the same bounds as that of Table \ref{LHS_bounds}, are constructed for validation/testing purposes for each of the six species.

\section{Constrained Dual Prediction of Mean and Standard Deviation Using FFNN With NLPD Loss Function}
\begin{figure}[htbp]
\centering
\includegraphics[width = .3\linewidth]{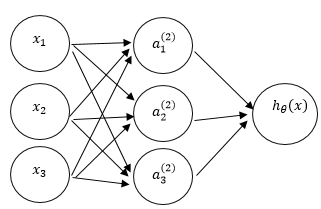}
\caption{Neural network mapping for a dummy 3-layer case with a three dimensional input and a univariate output.}
\label{mapping_nn}
\end{figure}

One of the most popular supervised machine learning techniques roughly based on the working of a human brain is the feed-forward neural network (FFNN). It is characterized by an input layer, an output layer, and layers in-between called hidden layers, where each layer is composed of \textit{neurons} or \textit{nodes}. An FFNN, especially one with multiple hidden layers (the so-called feed-forward deep neural network or FFDNN), often consists of a large number of parameters (\textit{weights} and \textit{biases}) that control the function mapping from one layer to the next. The depiction of the mapping for a dummy 3-layer FFNN is shown in Fig. \ref{mapping_nn}. The parameters $a_i^{(j)}$ in Fig. \ref{mapping_nn} are called \textit{activation units}, and their expressions are given in Eqs. \ref{a12}-\ref{a13}.   The functions $g_1$ and $g_2$ appearing in Eqs. \ref{a12}-\ref{a13} are some user-defined \textit{activation functions}, and $\tilde{\theta}^{(j)}$ denotes the matrix of weights controlling function mapping from layer $j$ to layer $(j+1)$. The elements of matrix $\tilde{\theta}^{(j)}$ are determined from the optimization of a user-defined $\tilde{\theta}^{(j)}$-dependent cost function. For more details on FFNN, see \cite{SVOZIL199743}. 

\begin{eqnarray}\label{a12}
a_1^{(2)}&=&g_1(\tilde{\theta}_{10}^{(1)}+\tilde{\theta}_{11}^{(1)}x_1+\tilde{\theta}_{12}^{(1)}x_2+\tilde{\theta}_{13}^{(1)}x_3)\\\label{a22}
a_2^{(2)}&=&g_1(\tilde{\theta}_{20}^{(1)}+\tilde{\theta}_{21}^{(1)}x_1+\tilde{\theta}_{22}^{(1)}x_2+\tilde{\theta}_{23}^{(1)}x_3)\\\label{a32}
a_3^{(2)}&=&g_1(\tilde{\theta}_{30}^{(1)}+\tilde{\theta}_{31}^{(1)}x_1+\tilde{\theta}_{32}^{(1)}x_2+\tilde{\theta}_{33}^{(1)}x_3)\\\label{a13}
a_1^{(3)}&=&h_{\theta}(x)=g_2(\tilde{\theta}_{10}^{(2)}+\tilde{\theta}_{11}^{(2)}a_1^{(2)}+\tilde{\theta}_{12}^{(2)}a_2^{(2)}+\tilde{\theta}_{13}^{(2)}a_3^{(2)})
\end{eqnarray}

In this paper, we use a variant of the FFNN that simultaneously predicts the mean and standard deviation of the drag coefficient as a function of the input features (see \cite{Licata2022}). For networks modeling the drag coefficient associated with the DRIA GSI model, we have the following eight features: $v_{\infty}$, $T_w$, $T_{\infty}$, $\alpha$, $\sin{\beta}$, $\cos{\beta}$, $\sin{\Phi}$, $\cos{\Phi}$. Instead of a univariate output, we predict an output with shape $[2, 1]$. The first output node represents the mean drag coefficient and the second output node represents the corresponding standard deviation. For the proposed variant of the FFNN, the input training data has an array structure of [number of samples, feature dimension], where the feature dimension is eight for DRIA. The output training data is augmented with zeros to have an array structure of $[$number of samples, 2, 1$]$. Since standard deviation values can only be non-negative, we constrain the second output node to be non-negative using the \textit{softplus function} so that meaningful uncertainties are predicted. One may also use alternatives such as the \textit{absolute value function} to impose the positivity constraint on the standard deviation prediction, but it may lead to unstable predictions because of the non-smooth first derivative at zero. For training purposes, we use the \textit{negative logarithm of the probability density} (NLPD) loss function, given as:

\begin{equation}\label{NLPD}
loss_{NLPD} = \frac{1}{n_t}\sum_{i=1}^{n_t}\Bigg[\log{\hat{\sigma_i}^2}+\frac{||\mathbf{y_i}-\mathbf{\hat{\mu}_{i}}||^2}{\hat{\sigma_i}^2} + \log{2\pi}\Bigg]
\end{equation}
where $(\mathbf{x_i,y_i})_{i=1,...,n_t}$ represent the training data set, $\mathbf{\hat{\mu}_i}$ represents the prediction mean, and $\hat{\sigma_i}^2$ represents the prediction variance. 

\begin{figure}
\centering
\includegraphics[width = .5\linewidth]{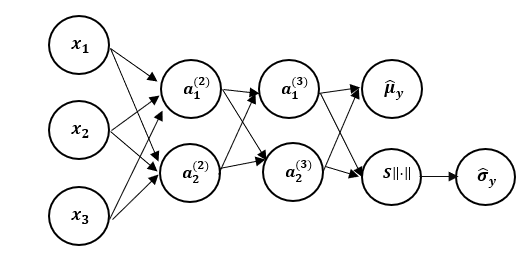}
\caption{Dummy feed-forward neural network for dual prediction of the mean and standard deviation. The non-negative constraint on the standard deviation prediction is indicated by the softplus function $S\| \cdot\| $.}
\label{joey_ffnn}
\end{figure}

Although the training data set only contains a cosmetic set of zeros as standard deviations, we will see later in this paper that the functional form of NLPD is powerful enough to allow the network to learn well-calibrated uncertainty estimates while also producing reasonably accurate mean estimates. The implementation of the model is carried out using the Keras deep learning API \citep{chollet2015keras}, where we define a custom class for modifying the in-built dense layers to meet our constraints. A dummy FFNN with three inputs for dual prediction of the mean and standard deviation of a variable $y$ is shown in Fig. \ref{joey_ffnn}. 

\section{Model Calibration}
Calibration is the requirement in stochastic modeling that the predicted probabilities approximate the probability of actual events \citep{CC2020}. A well-calibrated model (assuming Gaussian distribution), for example, should have around 68$\%$ of observations within one standard deviation of the predicted mean, 95$\%$ of observations within two standard deviations of the predicted mean, and 99.7$\%$ of observations within three standard deviations of the predicted mean. Uncalibrated models tend to be overconfident or underconfident in their predictions, and one should not trust their inferences. A convenient way to check how well a model is calibrated is by looking at its ``consistency curve" or ``calibration curve".  

Let the expected confidence interval levels be: $C =[5\%, 10\%, 15\%,....., 95\%]$. The corresponding coefficients defining the uncertainty bounds are then given as: $\zeta[k] = \sqrt{2}$      $\mathrm{ erf^{-1}}(C[k]/100)$, where erf is the well-known error function. Let $(\mathbf{x_{o_j},y_{o_j}})_{j=1,...,m}$ be the observation data set and let the corresponding predictions be $(\mathbf{{\mu}_{j}, \sigma_{j}})_{j=1,...,m}$, where $\mathbf{{\mu}_{j}}$ represents the mean and $\mathbf{\sigma_{j}}$ represents the standard deviation. Then, the percentage of the observed data set within the lower and upper uncertainty bounds associated with $C[k]$ is obtained as \citep{AGSBAT2020}: 

\begin{equation}
P[k] = \Bigg[\sum_{j=1}^{m}\frac{\mathcal{I}\big(({\mu}_{j}-\zeta[k]\sigma_{j})<y_{o_j}<({\mu}_{j}+\zeta[k]\sigma_{j})\big)}{m}\Bigg] \times 100
\end{equation}
where $\mathcal{I}$ is the indicator function. 

The consistency curve mentioned earlier is the plot of $P$ versus $C$. The proximity of the consistency curve to the $y=x$ line (i.e., a straight line with a slope of $45^0$ and passing through the origin) is used to measure calibration in this study. The consistency curve will perfectly overlap the $y=x$ line in a perfectly calibrated system (this does not usually happen in real-life models). 

In addition to the qualitative performance of the estimated uncertainties, we also investigate the quantitative performance using the mean absolute calibration error (MACE), defined as:

\begin{equation}
    MACE = \frac{1}{n_C}\sum_{k=1}^{n_C}|C[k]-P[k]|
\end{equation}
where $n_C$ is the number of confidence interval levels.

In this work, in order to improve the model calibration performance, we re-scale the predicted standard deviation values using the following scaling factor \citep{LIFKO2021}:

\begin{equation}\label{rescale}
s = \sqrt{\frac{1}{n_v}\sum_{i=1}^{n_v}\Bigg[\frac{||\mathbf{y_i}-\mathbf{\hat{\mu}_{i}}||^2}{\hat{\sigma_i}^2} \Bigg]}    
\end{equation}
where $n_v$ is the number of validation data samples, $\mathbf{y_i}$ is the TPMC drag coefficient, $\mathbf{\hat{\mu}_{i}}$ is the predicted mean drag coefficient, and $\hat{\sigma_i}^2$ is the predicted standard deviation for the $i^{th}$ validation sample.

\section{Prediction Performance for Different Data Sizes}
In this section, we first simulate drag coefficients for a sphere, assuming that the drag coefficient is attitude-independent. Following that, we carry out drag coefficient simulations for the CHAMP satellite. Because of its complex geometry and drag coefficient's dependence on attitude, we demonstrate that a satellite like the CHAMP necessitates a much larger number of data points for neural network training to achieve sufficient accuracy and reliability.

We perform neural network simulations for the spherical satellite using 10,000 data points for training/validation purposes for each of the species $H$, $He$, $N$, $N_2$, $O$, $O_2$. A critical task in the design of neural network architecture is to decide on the values of the model hyper-parameters. These are the parameters that we cannot learn from the training process. In our work, we restrict ourselves to the hyper-parameter optimization (also known as ``tuning") of the following quantities: (i) number of hidden layers, (ii) number of neurons in each hidden layer, (iii) activation functions, (iv) Monte Carlo dropout rates, (v) network optimizer, and (vi) batch size. The traditional practice uses a heuristic or manual approach to select the hyper-parameters. In our work, however, we use the KerasTuner library \citep{omalley2019kerastuner} to find our near-optimal hyper-parameters. Within KerasTuner, we have the following optimizer options: (i) random search, (ii) Bayesian optimization, and (iii) hyperband; this work uses Bayesian optimization. Table \ref{hyperparam_ffnnmself} lists the search space used for the hyper-parameter tuning and Table \ref{tuning_setup_ffnnmself} lists the essential tuning parameters.

\begin{table*}[htbp]
    \centering 
    \caption{Hyper-parameter search space.}
    \label{hyperparam_ffnnmself}
    \begin{tabular}{|l | l |} % Column formatting, 
    \hline
    Hyper-parameter & Values\\
    \hline
    Number of hidden layers & $[1,2,3,\cdots,9,10]$\\ \hline
    Number of neurons in each hidden layer & $[32,64,96,\cdots,512,1024]$\\ \hline
    Activation function for the neurons of each hidden layer &  [relu, tanh, sigmoid, softsign, selu, elu, linear] \\ \hline
    Dropout rate for each hidden layer & $[.05,.10,.15,\cdots,.75,.80]$\\ \hline
    Optimizer for neural network training & [rmsprop, adagrad, adam, nadam]\\ \hline
    Batch size & $[256,512,768,\cdots,3840,4096]$\\ \hline
   \end{tabular}
\end{table*}

\begin{table*}[htbp]
    \centering 
\caption{Tuning parameters used in the hyper-parameter optimization.}
\label{tuning_setup_ffnnmself}
    \begin{tabular}{|l | l |} % Column formatting, 
    \hline
Tuning Parameter & Value\\
\hline
Maximum number of trials & 150\\ \hline
Executions per trial & 5\\ \hline
Number of initial points & 50 \\ \hline
Early stop regularization patience & 50\\ \hline
Number of epochs & 200\\ \hline
   \end{tabular}
\end{table*}

\begin{figure*}
    \centering
    \subfloat[$C_D$ Prediction for $H$]{{\includegraphics[width=0.45\textwidth]{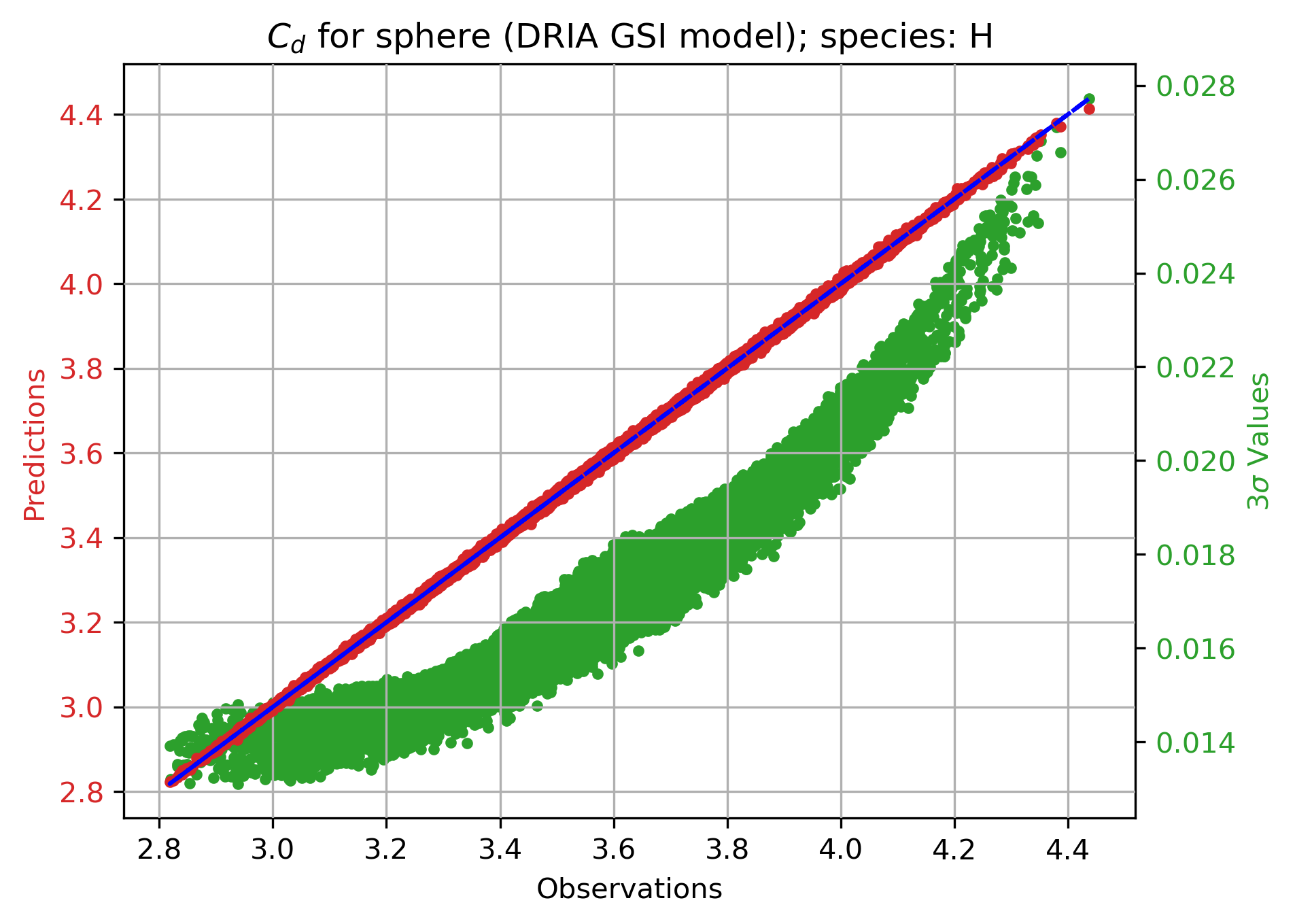}}}
    \subfloat[$C_D$ Prediction for $He$]{{\includegraphics[width=0.45\textwidth]{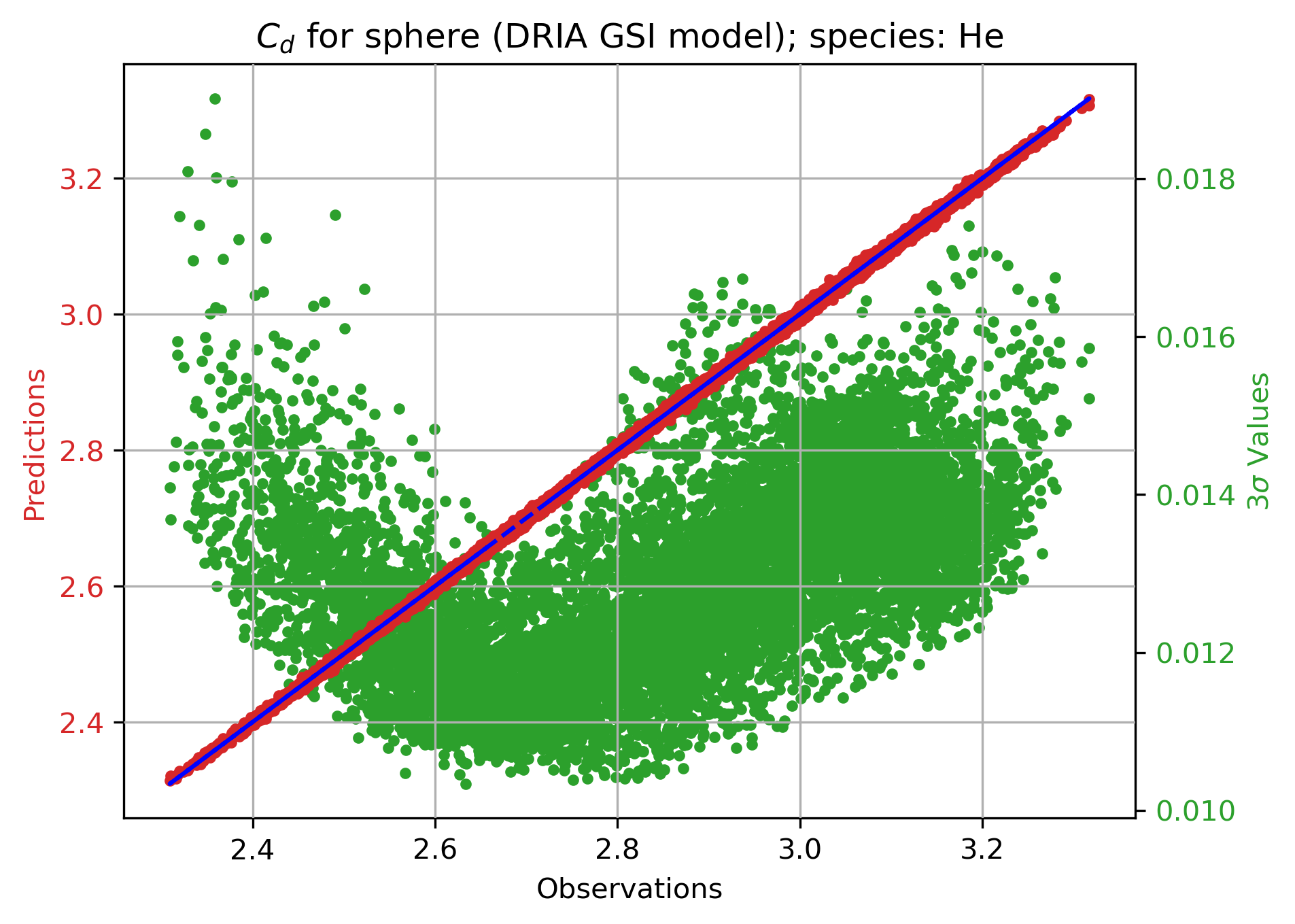} }}\\
    \subfloat[$C_D$ Prediction for $N$]{{\includegraphics[width=0.45\textwidth]{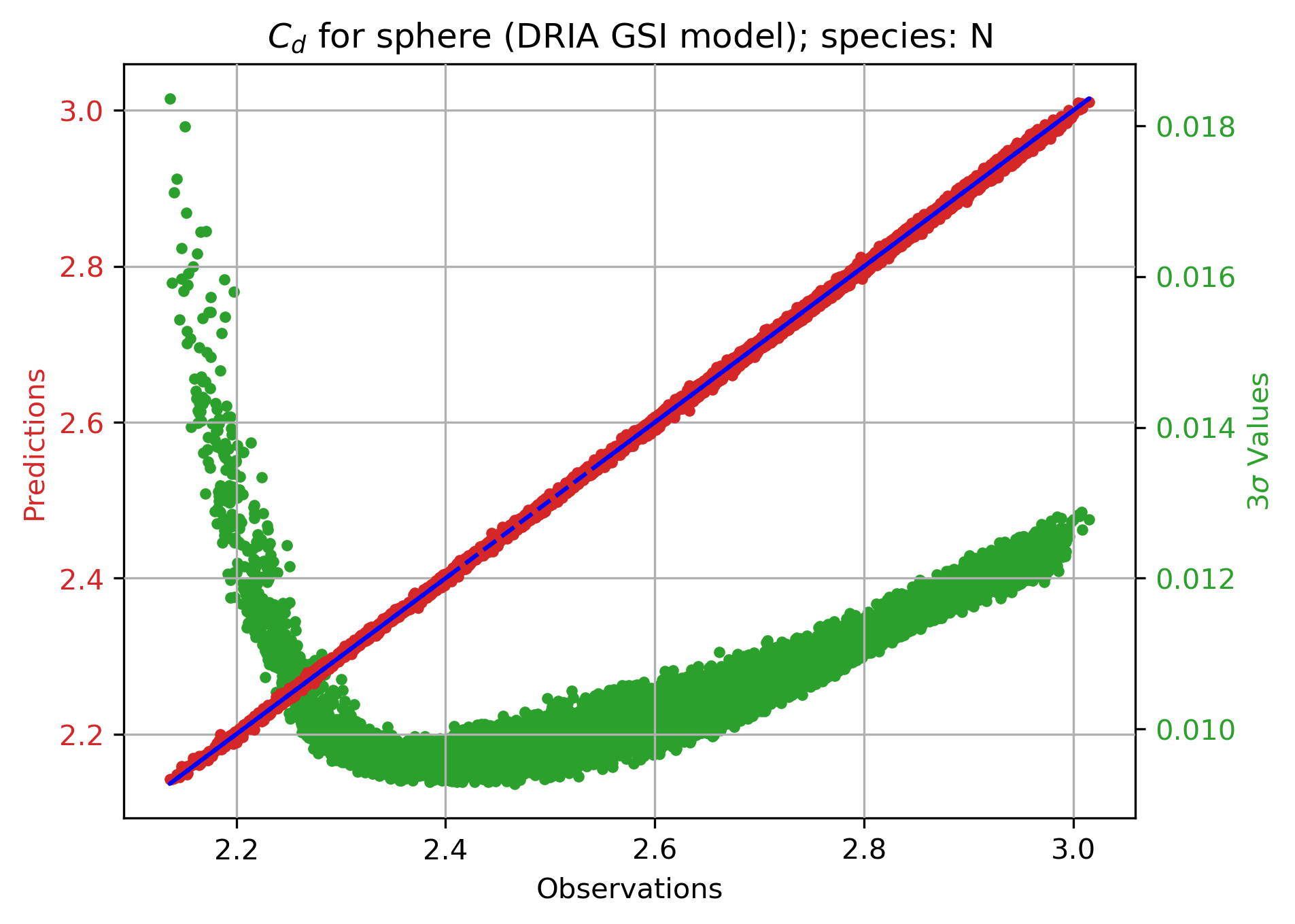} }}
    \subfloat[$C_D$ Prediction for $N_2$]{{\includegraphics[width=0.45\textwidth]{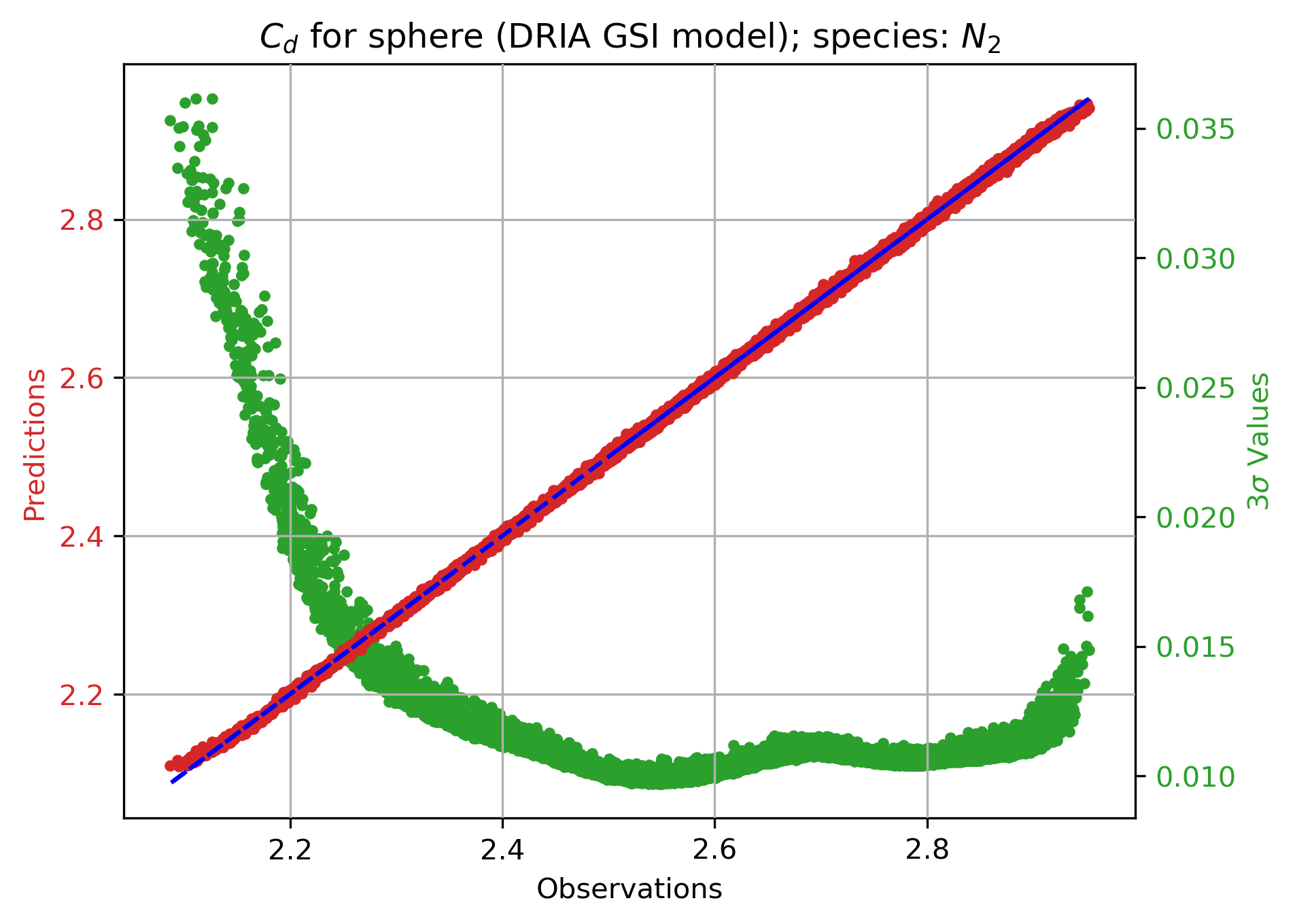} }}\\
    \subfloat[$C_D$ Prediction for $O$]{{\includegraphics[width=0.45\textwidth]{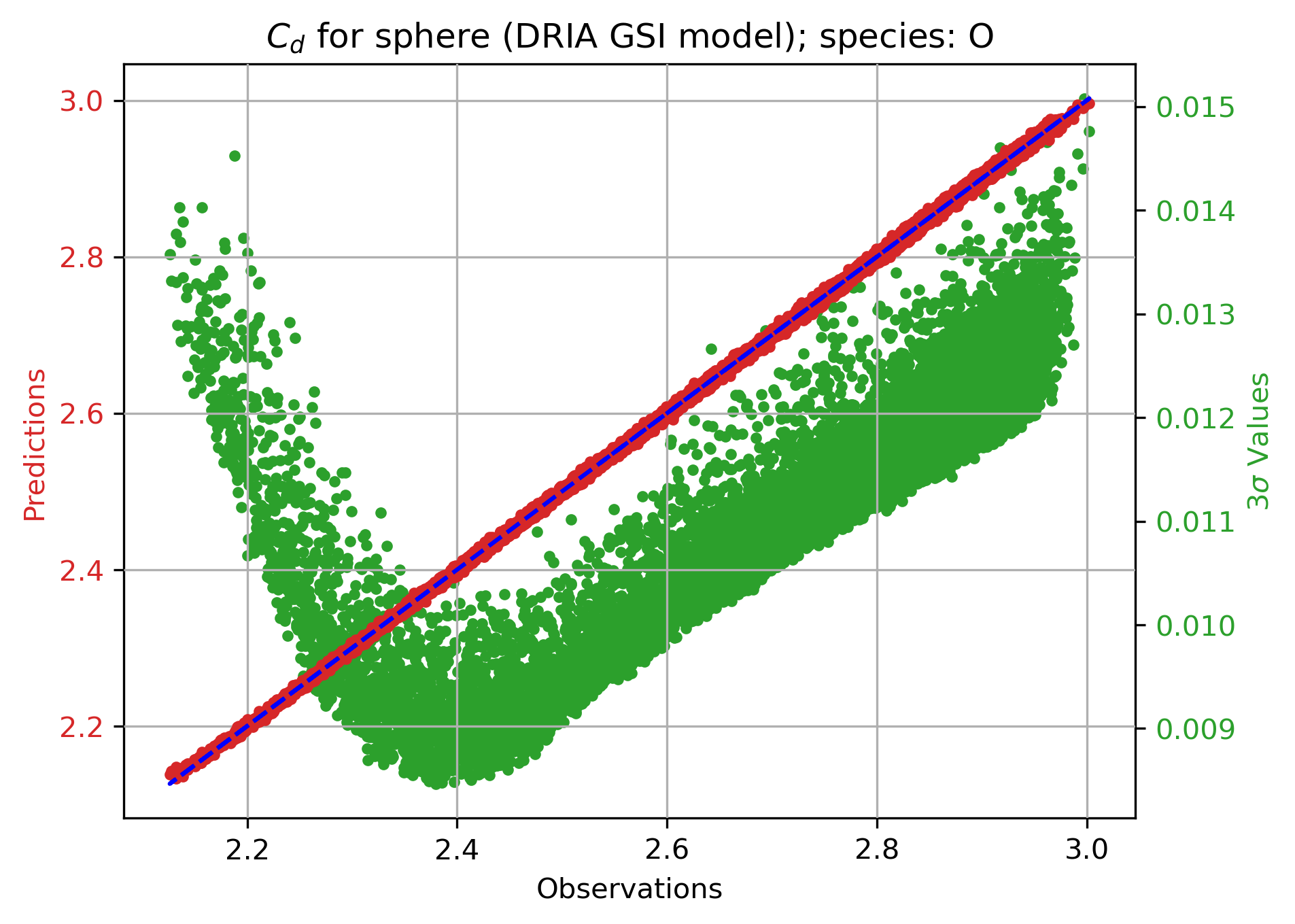} }} 
    \subfloat[$C_D$ Prediction for $O_2$]{{\includegraphics[width=0.45\textwidth]{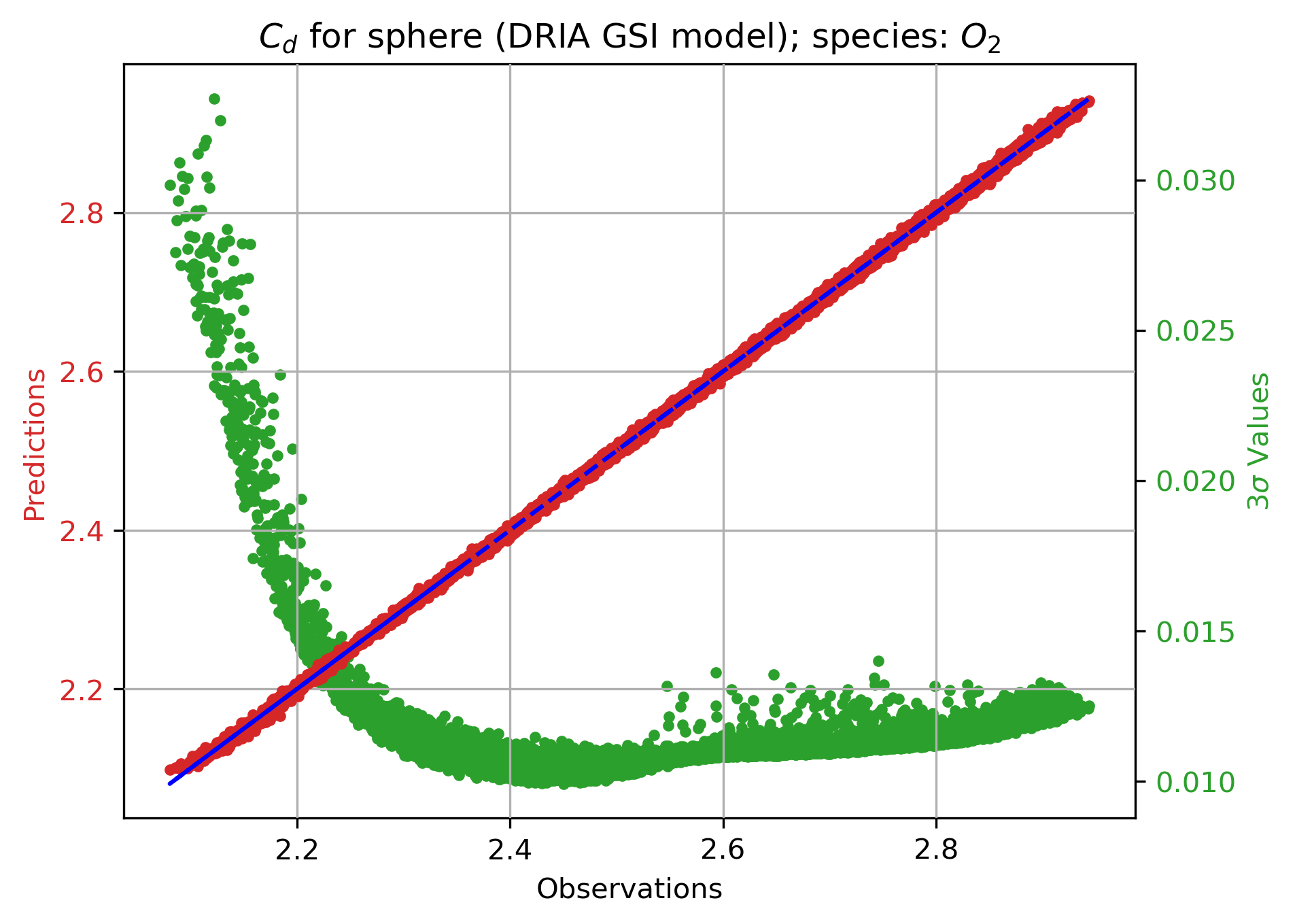} }}   
    \caption{Comparison of the true and predicted drag coefficients for the test data set for a spherical satellite. In red, we have the predicted means, and in green, we have the uncertainty predictions. Training is carried out using 8,500 samples. The testing is carried out using a different set of 10,000 samples.}
    \label{fig:10k_prediction_sphere}
\end{figure*}

We carry out the training using the best architecture resulting from the KerasTuner optimization process. The training data set comprises 8,500 samples obtained through a 15:85 split of 10,000 samples, with 15\% data samples used for validation and the remaining used for training. The performance of the trained network is assessed on a test data set of size 10,000. The 10,000 test samples are distinct from the training/validation samples and are generated using TPMC in the same way that the training data sets are. Note that the machine learning input dimension is now reduced to four (from eight for the CHAMP satellite) because the sphere is assumed to be rotationally invariant. Figure \ref{fig:10k_prediction_sphere} shows the neural network prediction results for the test data set. On the x-axis, we have the observed (numerical) drag coefficients; on the left y-axis, we have the predicted mean drag coefficients; on the right y-axis, we have the predicted 3$\sigma$ uncertainty values. The calibration curves corresponding to the prediction shown in Fig. \ref{fig:10k_prediction_sphere} are not included here for brevity. The RMSE and the MACE values for prediction on the test data set are given in Table \ref{rmsemace_allspecies_sphere}. For reference, the RMSE results obtained here are comparable to that of the results obtained by \cite{MEHTA20141590}, where the authors use Gaussian Process regression and 1000 ensemble points for training purposes. From Fig. \ref{fig:10k_prediction_sphere} and Table \ref{rmsemace_allspecies_sphere}, we can conclude that 10,000 data points are sufficient for producing accurate and well-calibrated results for a simple geometry like sphere.

\begin{table}[htbp]
    \centering 
    \caption{RMSE and MACE performance on the test data set for all species for a spherical satellite.}
   \label{rmsemace_allspecies_sphere}
   \begin{tabular}{|l | l | l |} % Column formatting, 
      \hline
      Species & RMSE & MACE (\%)\\
      \hline 
      H      &  0.0058 & 0.2879 \\
      \hline
      He      &  0.0044 & 0.4810 \\
      \hline
      N      &  0.0038 & 0.9447 \\
      \hline
      $N_2$      &  0.0038 & 0.5584 \\
      \hline
      O      &  0.0037 & 0.7421 \\
      \hline
      $O_2$      &  0.0038 & 0.8310 \\
      \hline
   \end{tabular}
\end{table}

 Next, we perform simulations for the CHAMP satellite. We perform a sensitivity analysis to determine the appropriate data size for neural network training. We inspect three different data sizes - (1) training with 10,0000 samples, (2) training with 20,000 samples, and (3) training with 50,000 samples for each of the species $H$, $He$, $N$, $N_2$, $O$, $O_2$. The performance of the trained networks is assessed on test data sets. In addition, the neural network predicted drag coefficients are compared to SPARTA-based drag coefficient data provided by Dr. Christian Siemes of the Delft University of Technology \citep{March2021}. The accuracy of the neural network predictions compared to the SPARTA-based drag coefficients and the computational costs determine the appropriate data size. For interested readers, Stochastic PArallel Rarefied-gas Time-accurate Analyzer (SPARTA) \citep{plimpton2019direct} is a parallel Direct Simulation Monte Carlo (DSMC) code for performing simulations of low-density gases in 2D or 3D.

\begin{figure*}
    \centering
    \subfloat[$C_D$ Prediction for $H$]{{\includegraphics[width=0.45\textwidth]{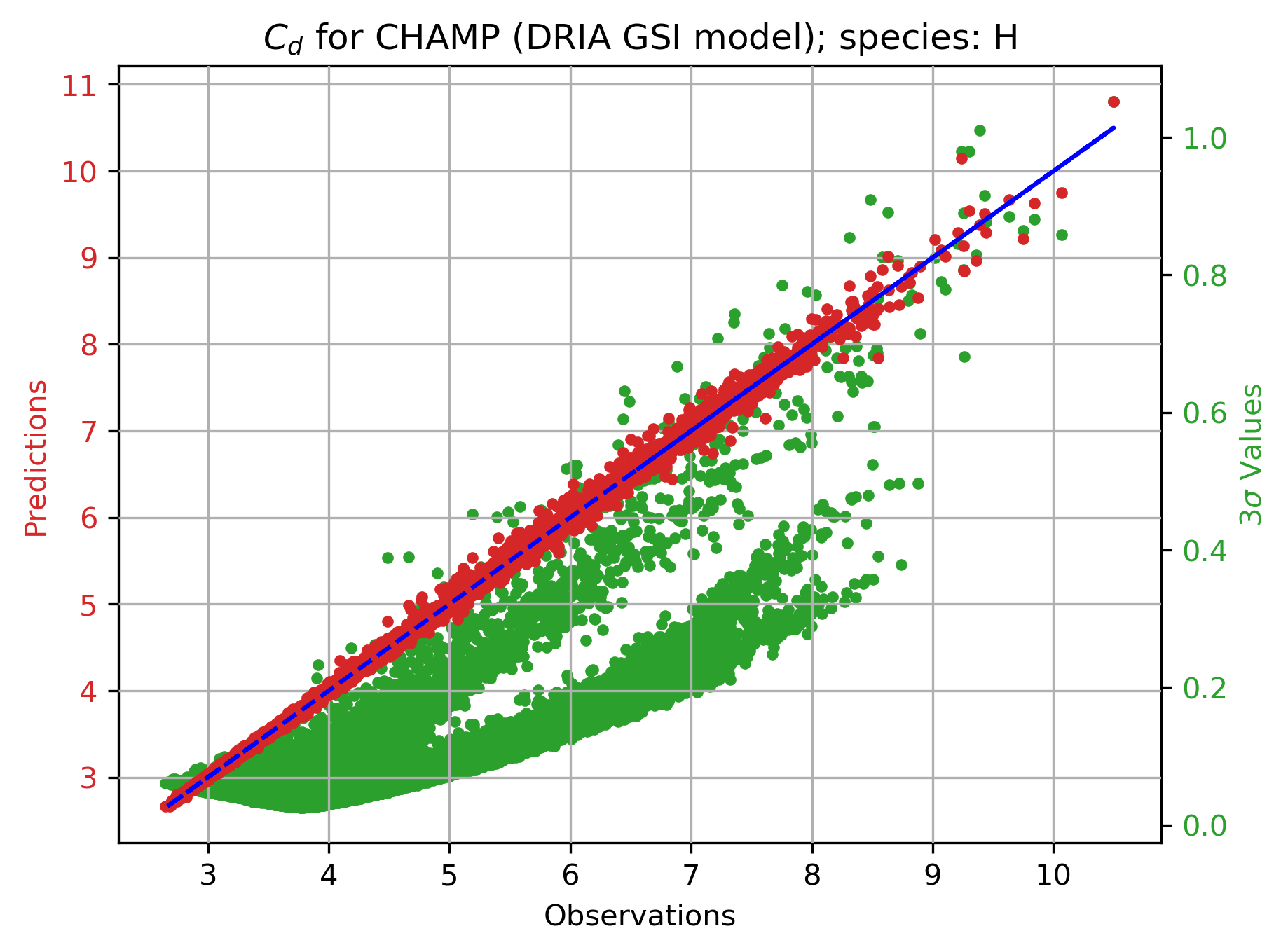}}}
    \subfloat[$C_D$ Prediction for $He$]{{\includegraphics[width=0.45\textwidth]{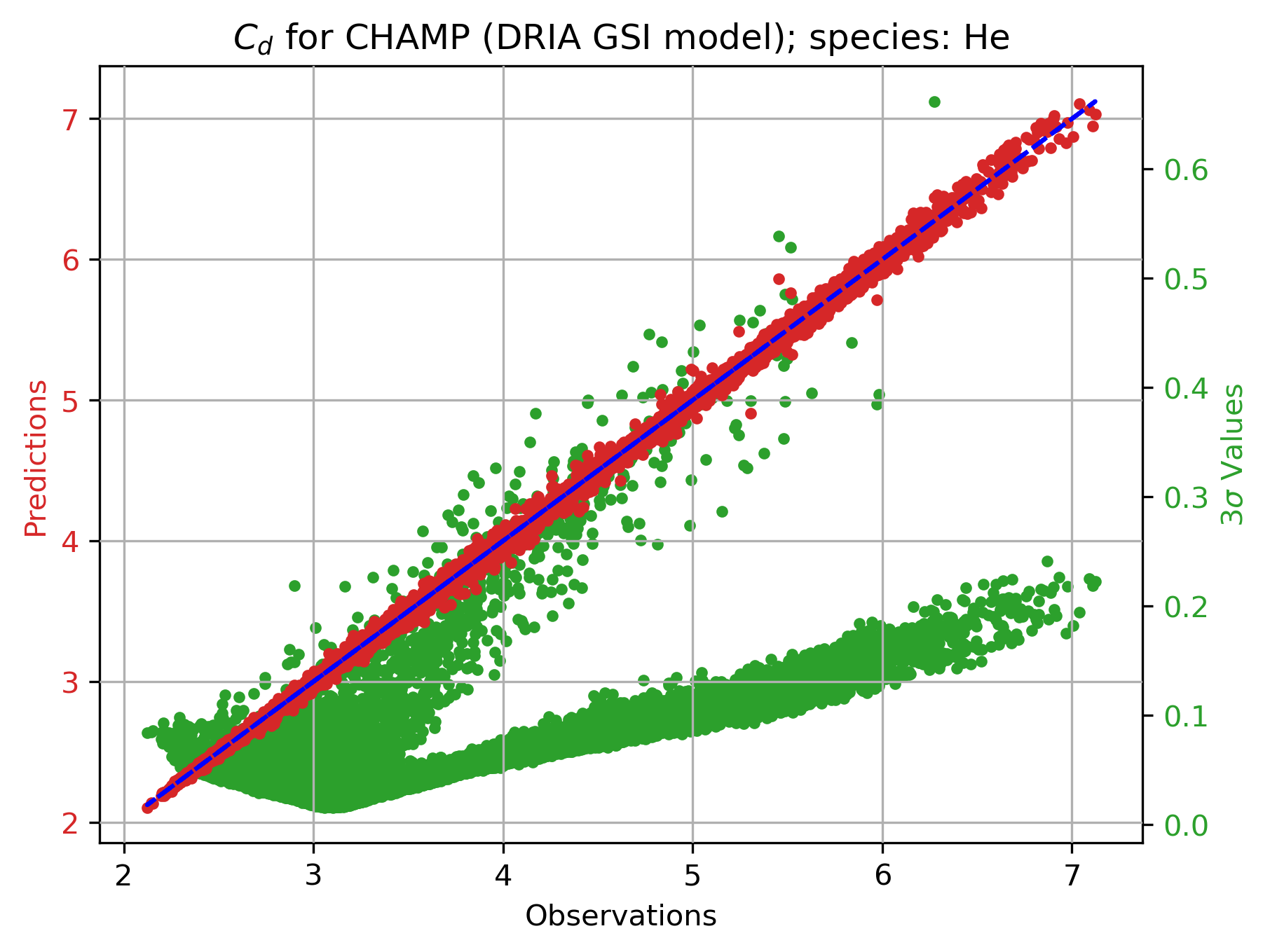} }}\\
    \subfloat[$C_D$ Prediction for $N$]{{\includegraphics[width=0.45\textwidth]{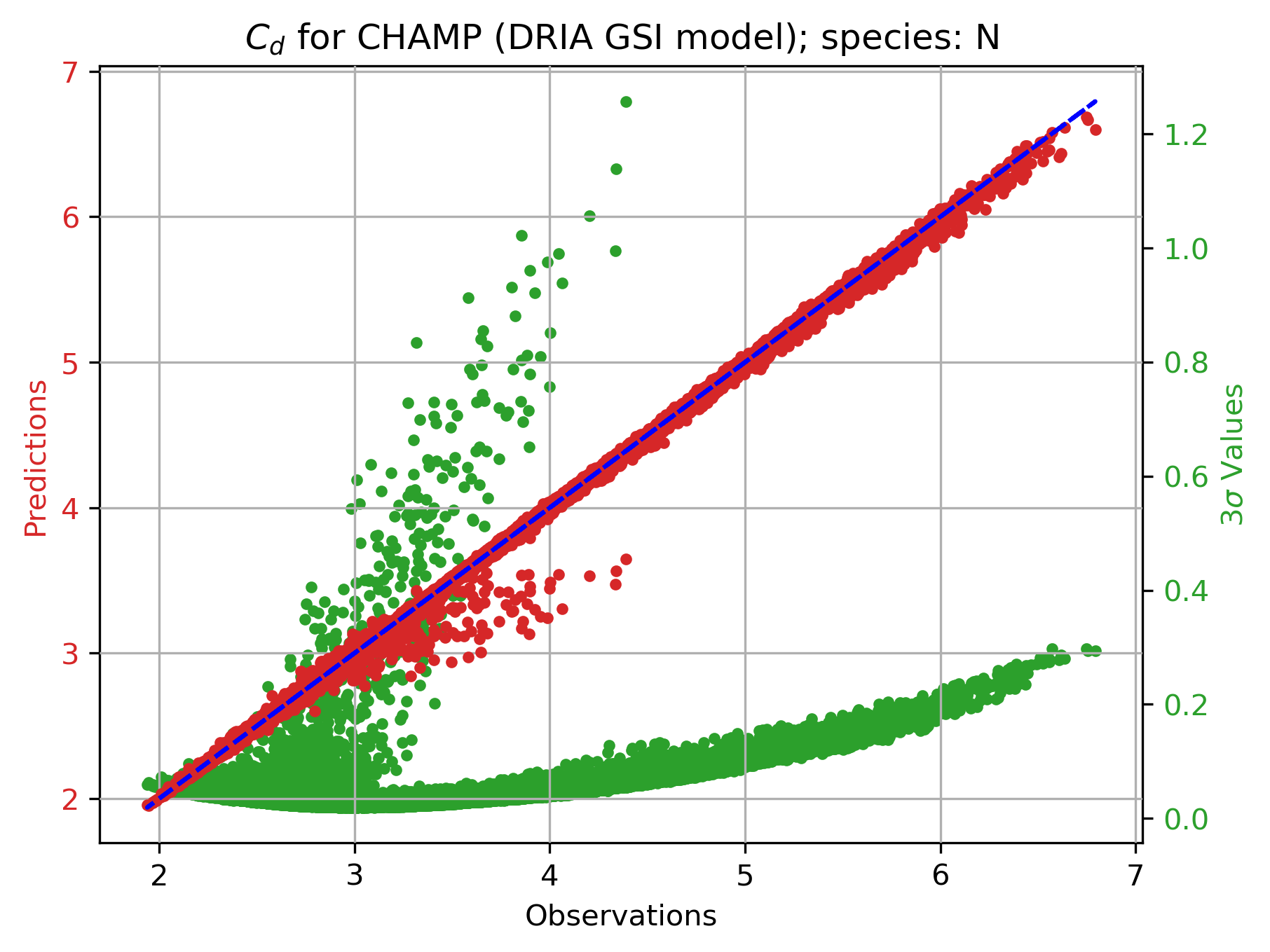} }}
    \subfloat[$C_D$ Prediction for $N_2$]{{\includegraphics[width=0.45\textwidth]{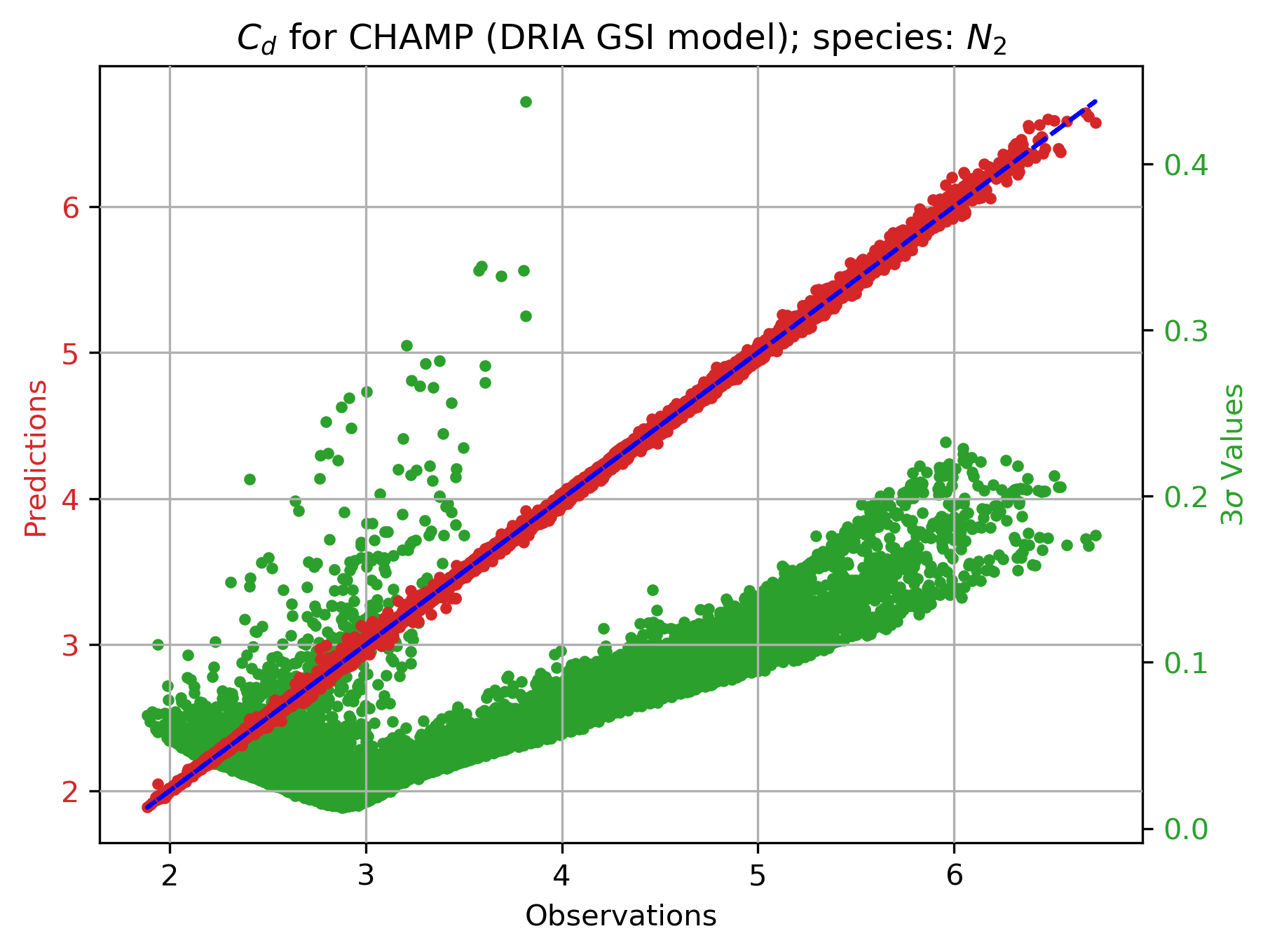} }}\\
    \subfloat[$C_D$ Prediction for $O$]{{\includegraphics[width=0.45\textwidth]{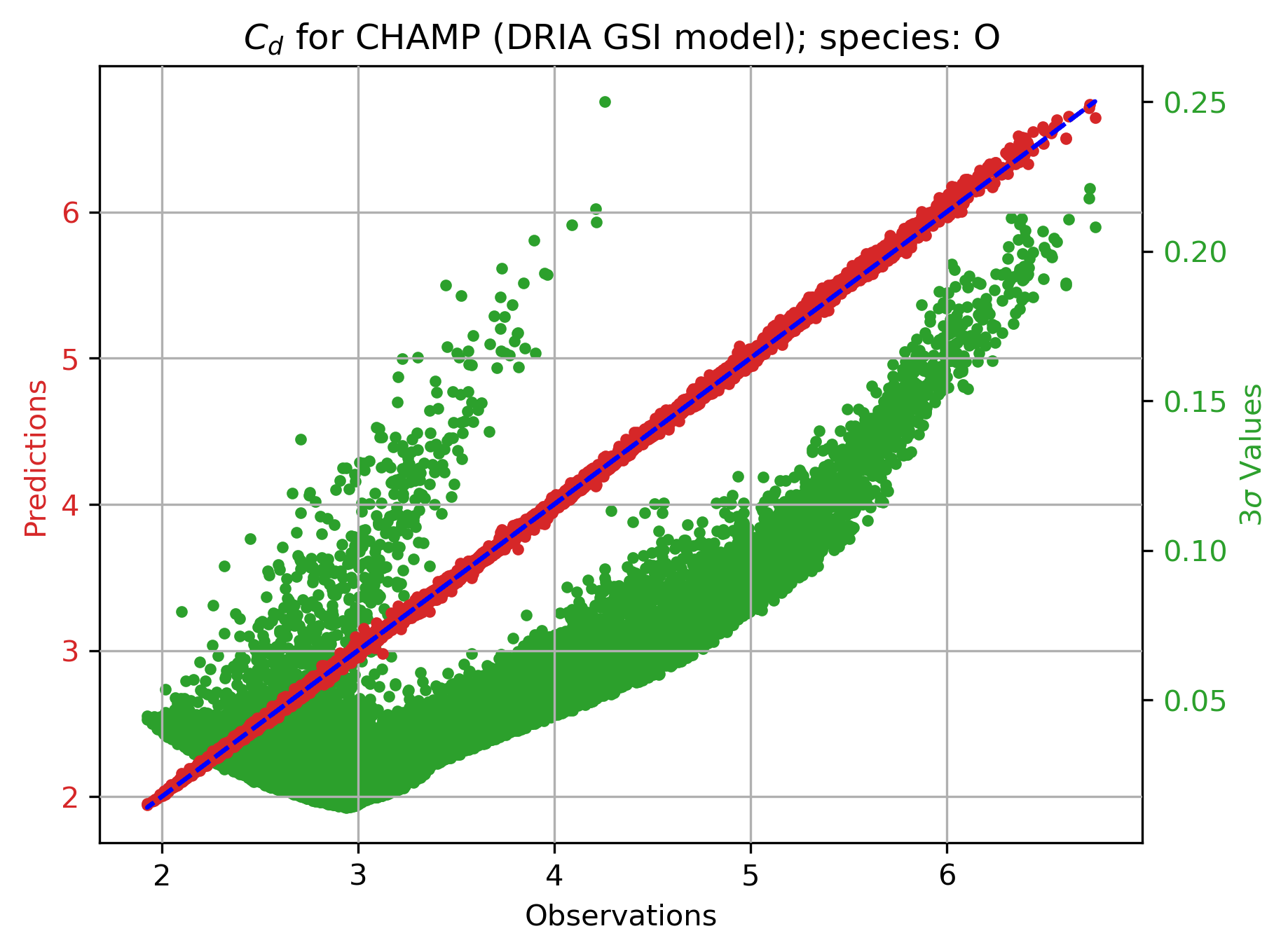} }} 
    \subfloat[$C_D$ Prediction for $O_2$]{{\includegraphics[width=0.45\textwidth]{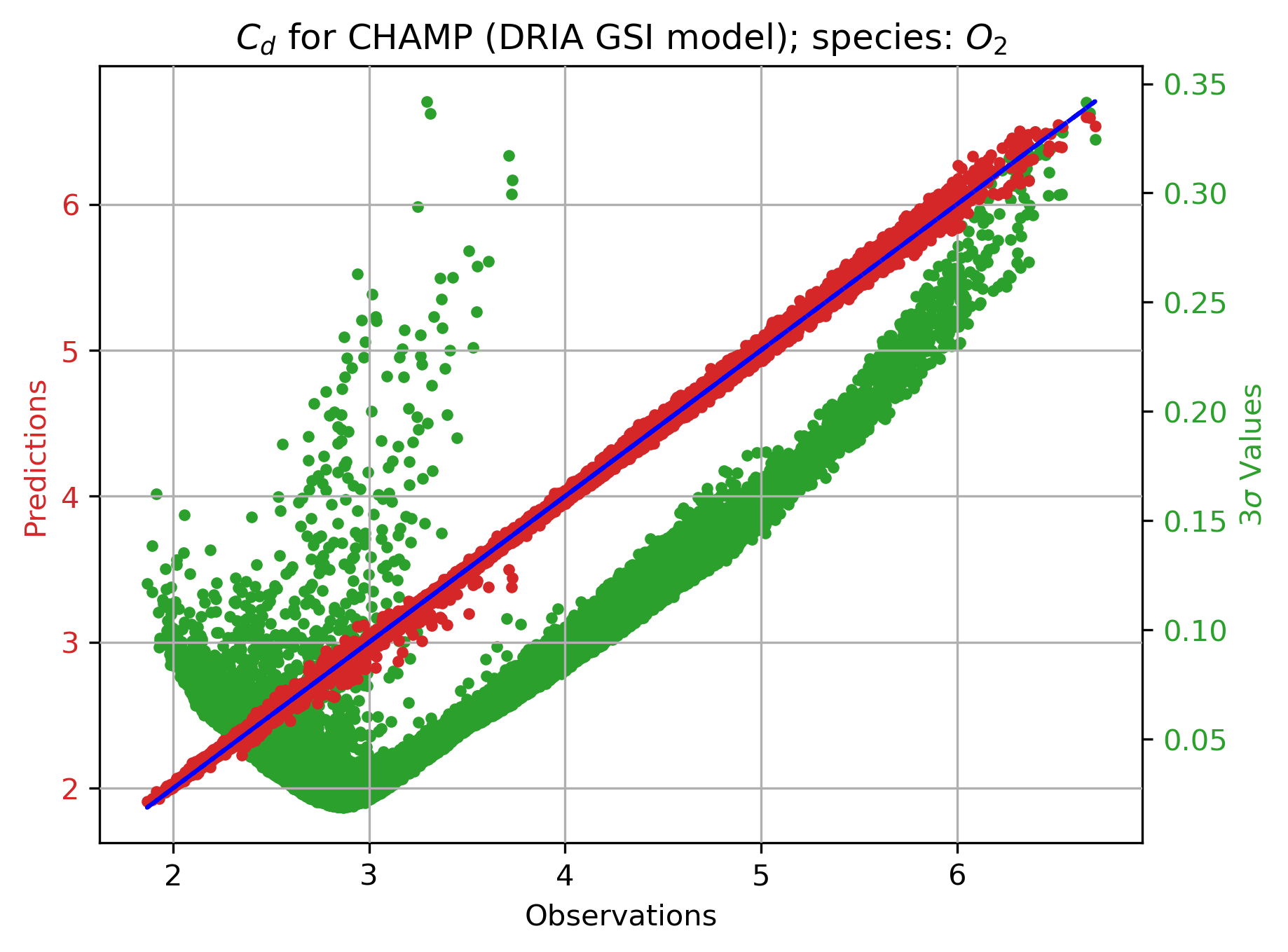} }}   
    \caption{Comparison of the true and predicted drag coefficients for the test data set for the CHAMP satellite. In red, we have the predicted means, and in green, we have the uncertainty predictions. Training is carried out using 50,000 samples. The testing is carried out using a different set of 42,500 samples.}
    \label{fig:50k_prediction}
\end{figure*}

\begin{figure*}
    \centering
    \subfloat[Calibration curve for $H$]{{\includegraphics[width=0.45\textwidth]{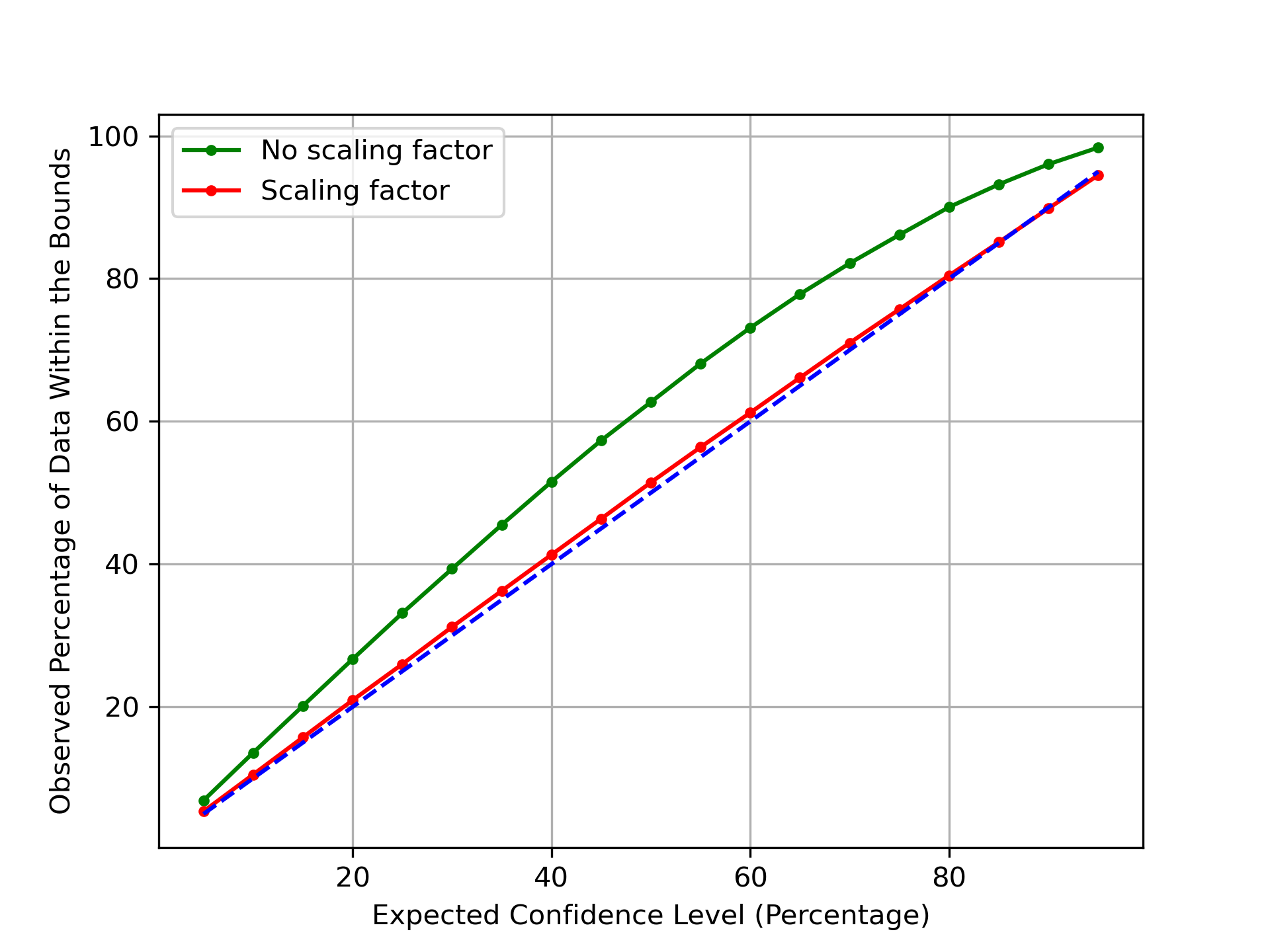}}}
    \subfloat[Calibration curve for $He$]{{\includegraphics[width=0.45\textwidth]{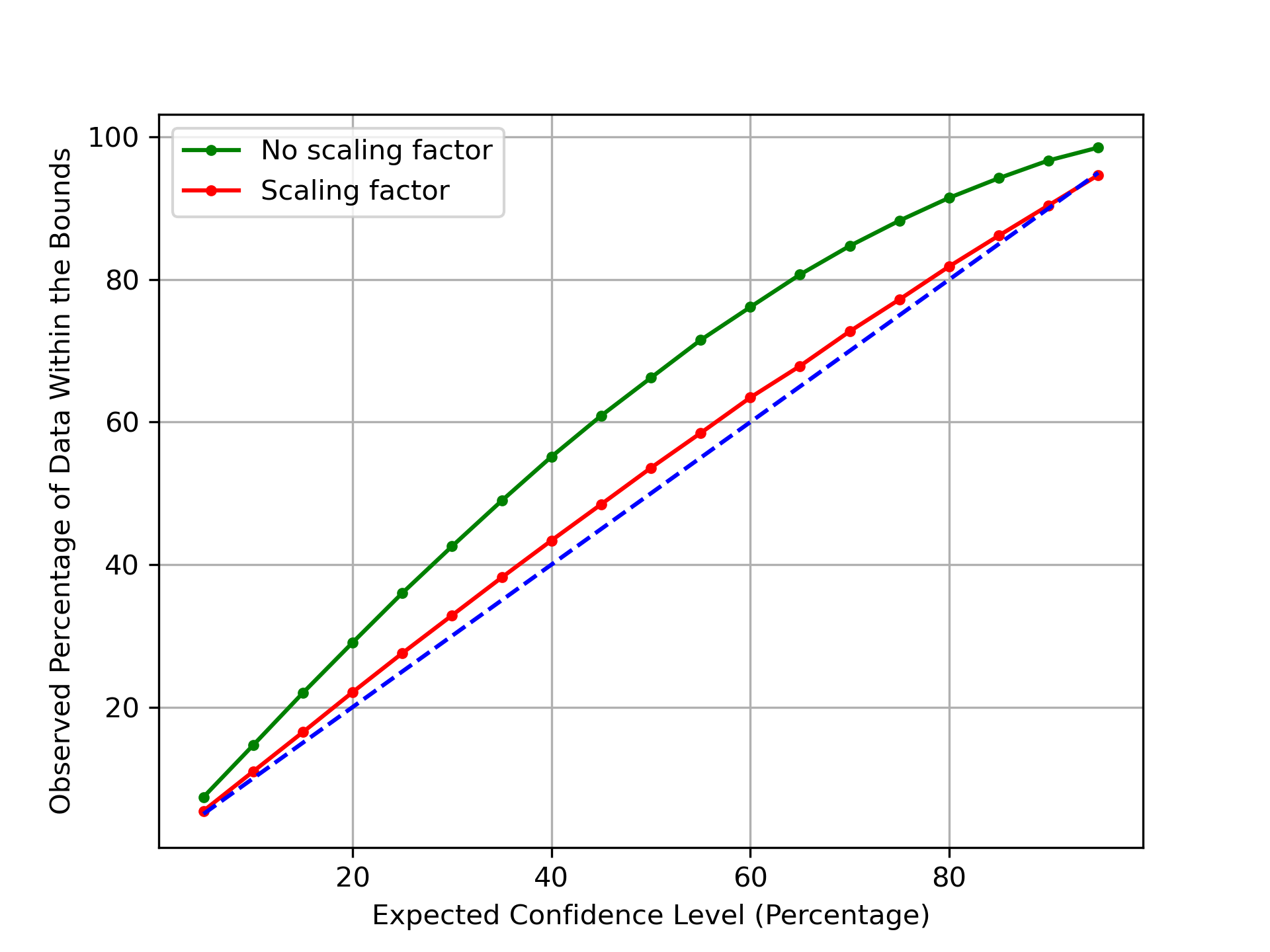} }}\\
    \subfloat[Calibration curve for $N$]{{\includegraphics[width=0.45\textwidth]{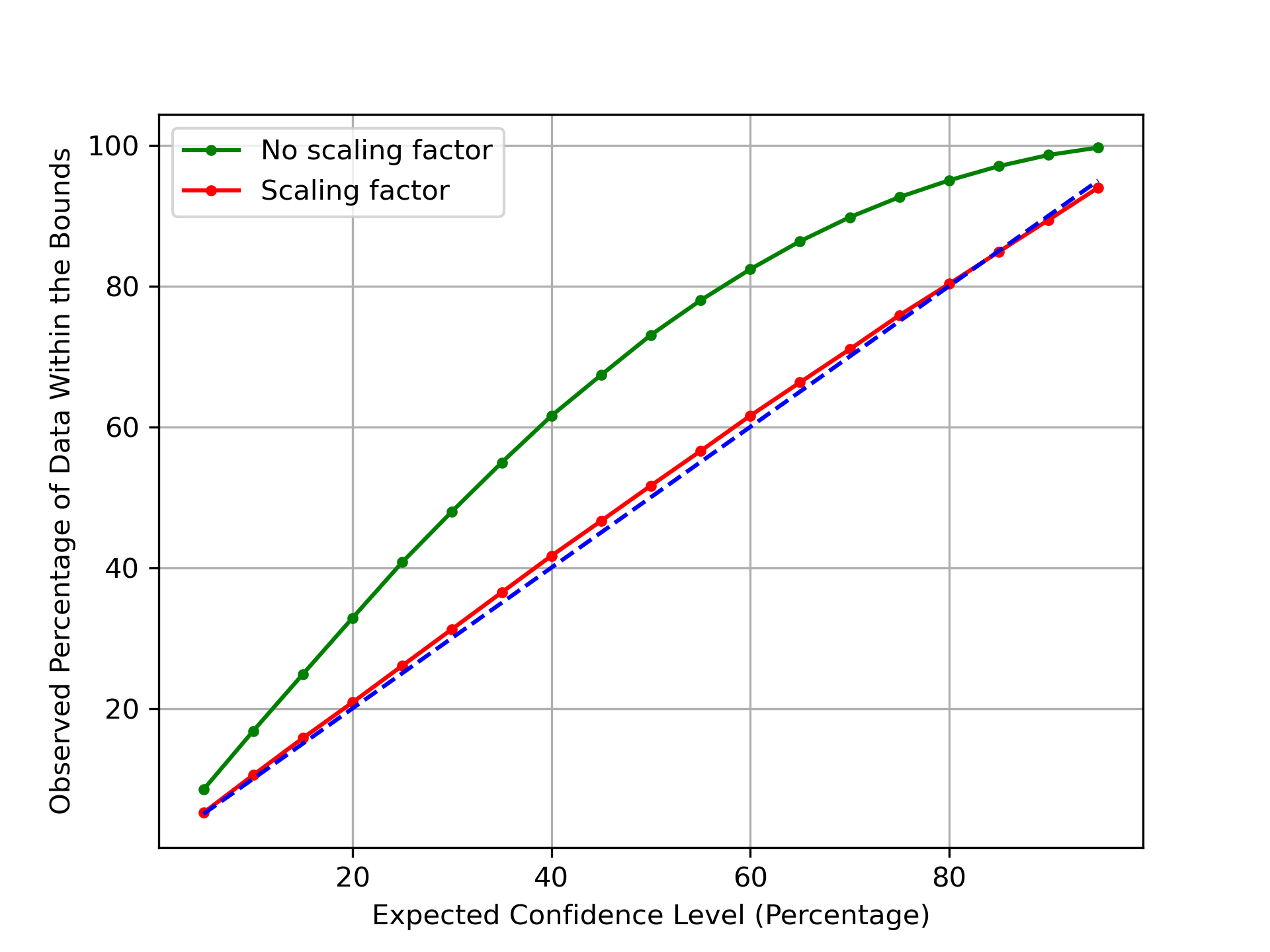} }}
    \subfloat[Calibration curve for $N_2$]{{\includegraphics[width=0.45\textwidth]{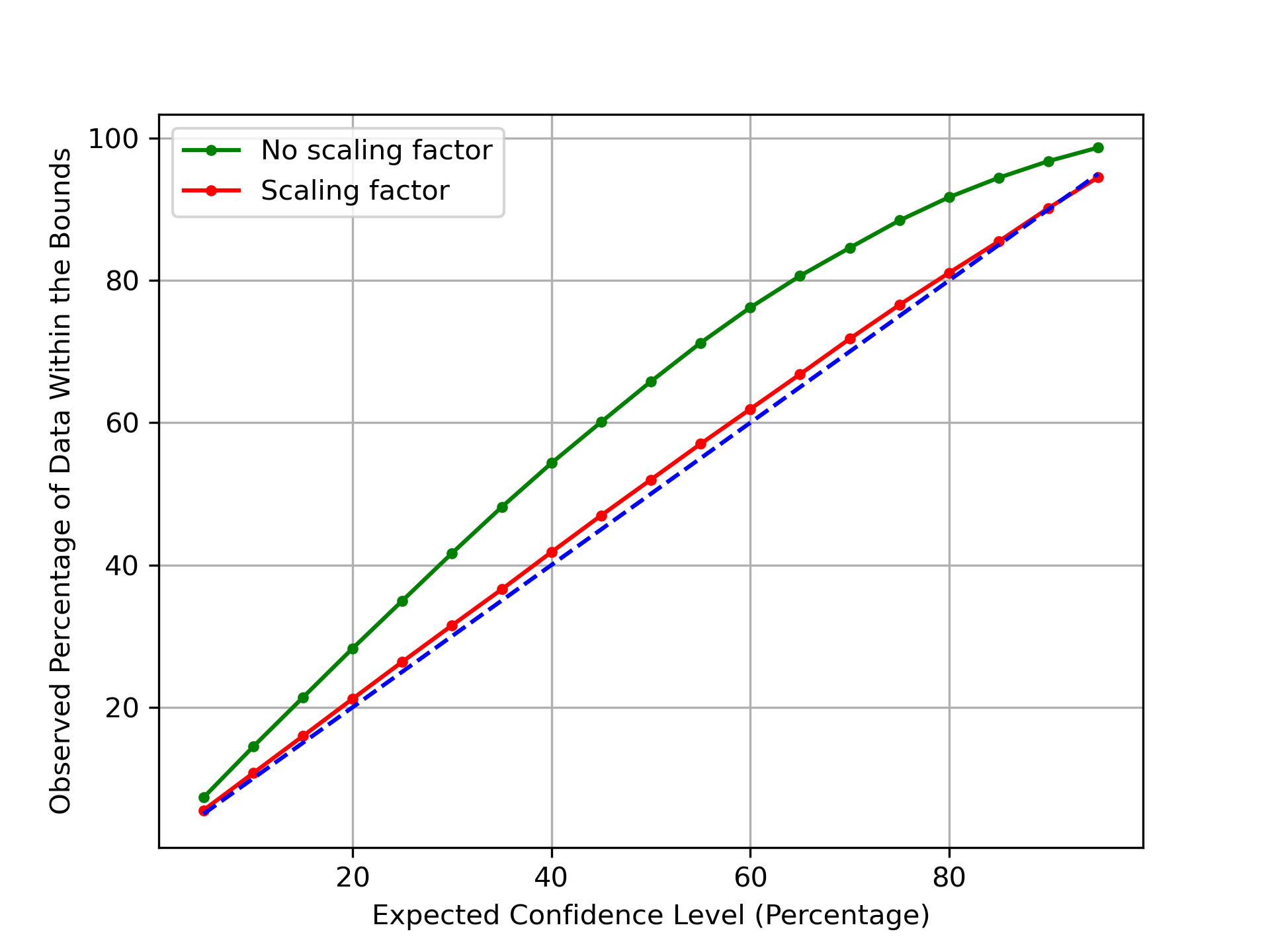} }}\\
    \subfloat[Calibration curve for $O$]{{\includegraphics[width=0.45\textwidth]{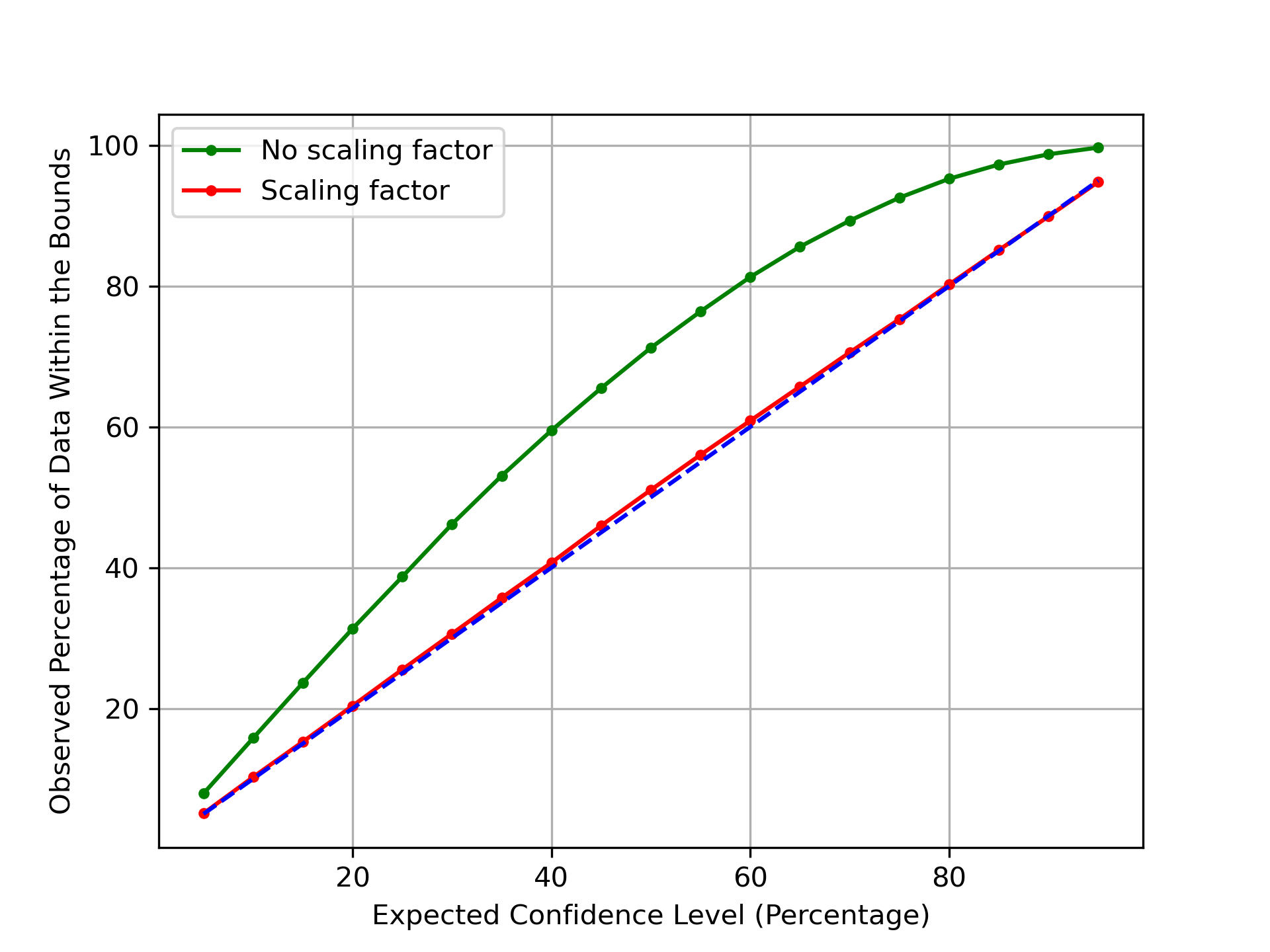} }} 
    \subfloat[Calibration curve for $O_2$]{{\includegraphics[width=0.45\textwidth]{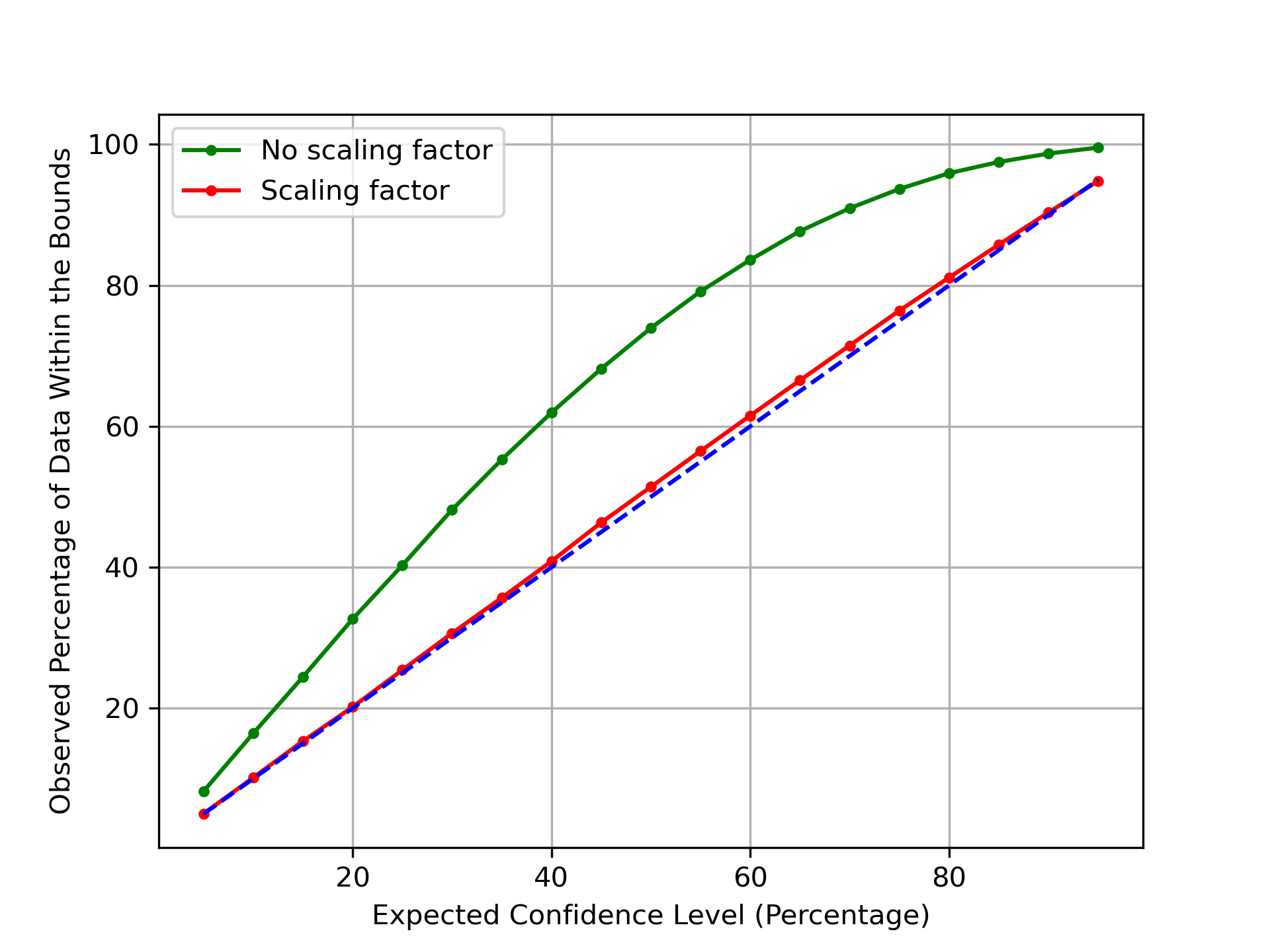} }}   
    \caption{Calibration curves for the predicted drag coefficient for the test data set for the CHAMP satellite. The prediction models are trained using 50,000 samples.}
    \label{fig:50k_prediction_calibration}
\end{figure*}

We carry out the training for the three data sizes (10,000, 20,000, and 50,000) using the best architecture resulting from the KerasTuner optimization process (based on Tables \ref{hyperparam_ffnnmself} and \ref{tuning_setup_ffnnmself}). Subsequently, we use the trained networks for drag coefficient prediction for the test data set. The test data set contains 42,500 samples obtained through a 15:85 split of 50,000 samples, with 15\% data samples used for validation and the remaining 85\% data samples used for testing. The 50,000 validation/test samples are distinct from the training samples. For training using 50,000 samples, the prediction performance on the test data set is shown in Fig. \ref{fig:50k_prediction}. The observed drag coefficients are on the x-axis, the neural network predicted mean drag coefficients are on the left y-axis, and the neural network predicted $3\sigma$ uncertainty values are on the right y-axis. The calibration performance for prediction on the test data set for models trained using 50,000 samples is given in Fig. \ref{fig:50k_prediction_calibration}. We have the calibration curve in red, and in blue, we have the reference $y=x$ line. Also shown in the figure, in green, we have the calibration curve corresponding to the case if we were ``not'' to re-scale the neural network predicted standard deviations using Eq. \ref{rescale}. It is evident from Fig. \ref{fig:50k_prediction_calibration} that the neural network models provide well-calibrated estimates. We do not show prediction performance plots and calibration curves for training using 10,000 and 20,000 samples for brevity. However, for completion, the RMSE and MACE values for all the species for training using 10,000, 20,000, and 50,000 samples are shown in Table \ref{rmsemace_allspecies}.

\begin{table*}[htbp]
    \centering 
    \caption{RMSE and MACE performance on the test data set for all species for the CHAMP satellite. Comparison is shown for training carried out using 10K samples, 20K samples, and 50K samples.}
   \label{rmsemace_allspecies}
   \begin{tabular}{|l | l | l |} % Column formatting, 
      \hline
      Species & RMSE (10K $|$ 20K $|$ 50K) & MACE (\%) (10K $|$ 20K $|$ 50K)\\
      \hline 
      H      &  0.0768 $|$ 0.0686 $|$ 0.0295 & 0.9511 $|$ 0.7778 $|$ 0.8845 \\
      \hline
      He      &  0.0337 $|$ 0.0485 $|$ 0.0188 & 3.3381 $|$ 0.2550 $|$ 2.2404 \\
      \hline
      N      &  0.0248 $|$ 0.0210 $|$ 0.0250 & 2.4136 $|$ 1.5145 $|$ 1.0279 \\
      \hline
      $N_2$      &  0.0368 $|$ 0.0251 $|$ 0.0159 & 0.6852 $|$ 0.4904 $|$ 1.3170 \\
      \hline
      O      &  0.0634 $|$ 0.0475 $|$ 0.0142 & 1.9360 $|$ 1.6784 $|$ 0.5117 \\
      \hline
      $O_2$      &  0.0266 $|$ 0.0445 $|$ 0.0209 & 2.2379 $|$ 1.6260 $|$ 0.8384 \\
      \hline
   \end{tabular}
\end{table*}

Next, for comparison of the neural network-predicted drag coefficients with the DSMC-predicted drag coefficients, the drag coefficients of the individual species are combined to compute the total drag coefficient. The total drag coefficient is given as \citep{MWK2014, WMK2014}:

\begin{equation}\label{CD}
C_D = f_{sc}C_{D_{ads}} + (1-f_{sc})C_{D_{surf}}
\end{equation}\normalsize
where $C_{D_{ads}}$ is the total drag coefficient based on a satellite completely covered by the adsorbate (atomic oxygen), and $C_{D_{surf}}$ is the total drag coefficient based on a clean satellite surface. The weight $f_{sc}$ is given as \citep{WMK2014}:

\begin{equation}
f_{sc} = \frac{K_{DRIA}P_o}{1+K_{DRIA}P_o}
\end{equation}
where $K_{DRIA}$ is the Langmuir adsorbate constant for the DRIA model (= $1.44 \times 10^6$) and $P_O$ is the partial pressure of atomic oxygen. The adsorbate and the surface drag coefficients are obtained from the drag coefficients of constituent species ($H$, $He$, $N$, $N_2$, $O$, $O_2$) using \citep{WMK2014}:

\begin{equation}
C_{D_{ads/surf}}=\bigg(\frac{1}{\sum_{k=1}^6 (\chi_k m_k)}\bigg)\sum_{k=1}^6 (\chi_k m_k C_{D_{ads/surf},k})
\end{equation}
where $\chi_k$ is the mole fraction of species $k$, $m_k$ is the mass of species $k$, and $C_{D_{ads/surf},k}$ is the drag coefficient for species $k$. The adsorbate drag coefficient corresponding to species $k$, i.e., $C_{D_{ads},k}$, is obtained by sampling from the distribution predicted by the neural network model for each species with features: [$v_{\infty}$, 400 K, $T_{\infty}$, $\alpha_{ads}$, $\sin{\beta}$, $\cos{\beta}$, $\sin{\Phi}$, $\cos{\Phi}$]. Similarly, the surface drag coefficient corresponding to species $k$, i.e., $C_{D_{surf},k}$, is obtained by sampling from the distribution predicted by the neural network model for each species with features: [$v_{\infty}$, 400 K, $T_{\infty}$, $\alpha_{surf}$, $\sin{\beta}$, $\cos{\beta}$, $\sin{\Phi}$, $\cos{\Phi}$]. The atmospheric translational temperature, $T_{\infty}$, is obtained from the NRLMSISE-00 model. We get the satellite relative velocity $v_{\infty}$ and the species mole fraction $\chi_k$ data from DSMC files provided by Dr. Christian Siemes. The energy accommodation coefficients $\alpha_{ads}$ and $\alpha_{surf}$ are taken as 0.85 because the DSMC results are generated using an accommodation coefficient of 0.85.

\begin{figure*}[ht]
    \centering
    \subfloat[24-hour prediction for March 27, 2002]{{\includegraphics[width=0.45\textwidth]{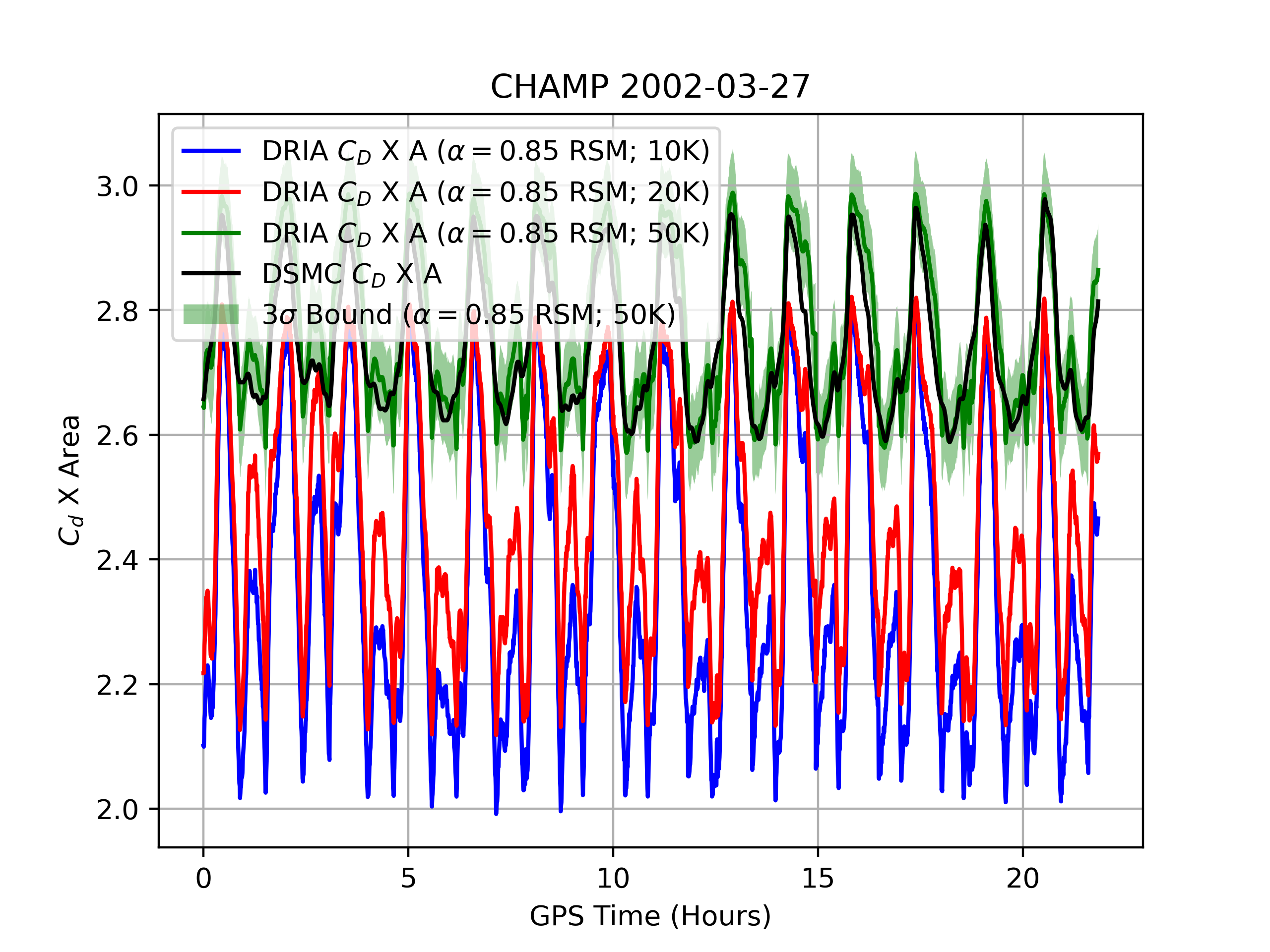}}}
    \subfloat[24-hour prediction for April 20, 2010]{{\includegraphics[width=0.45\textwidth]{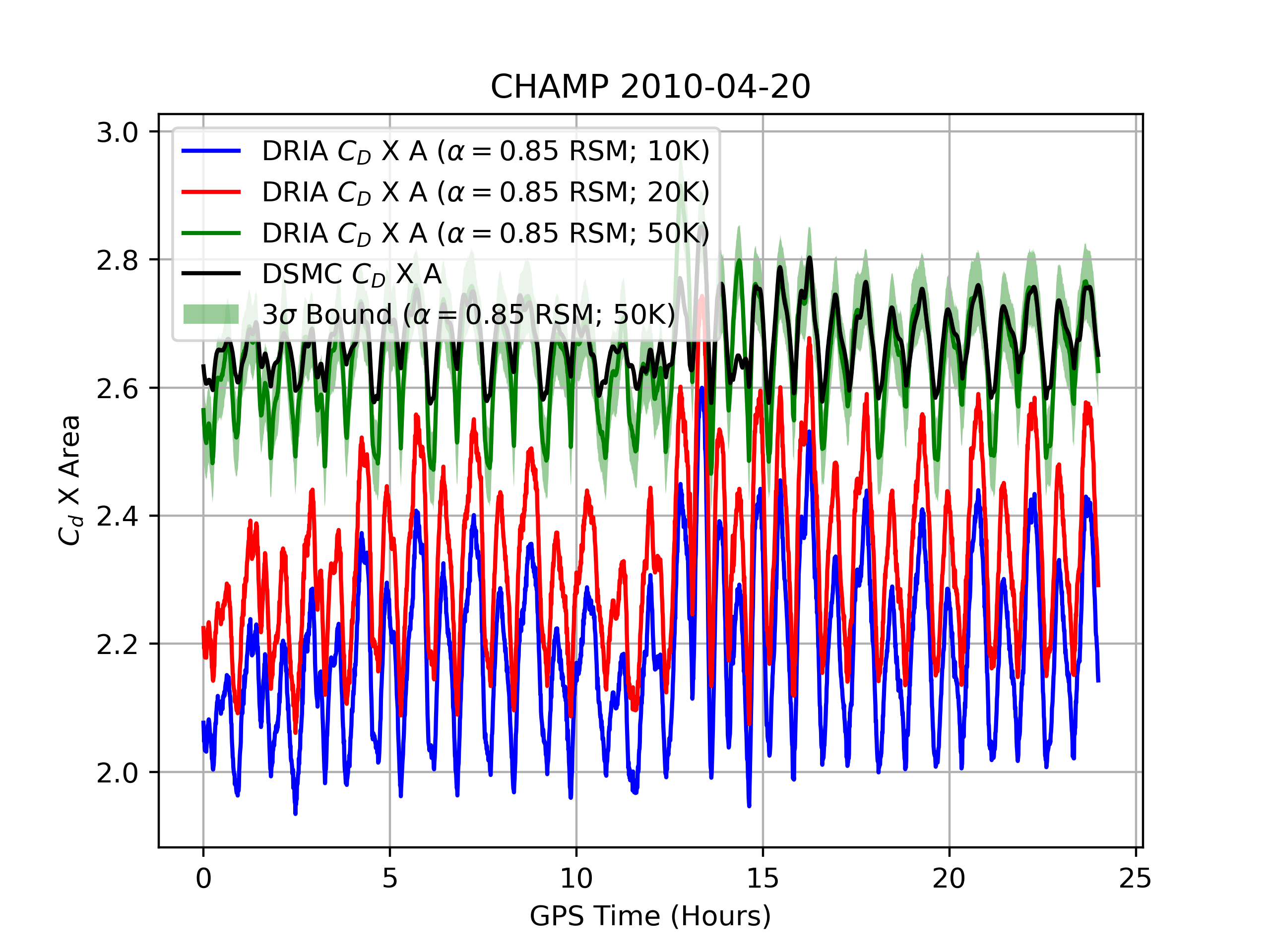} }}\\
    \subfloat[24-hour prediction for May 07, 2010]{{\includegraphics[width=0.45\textwidth]{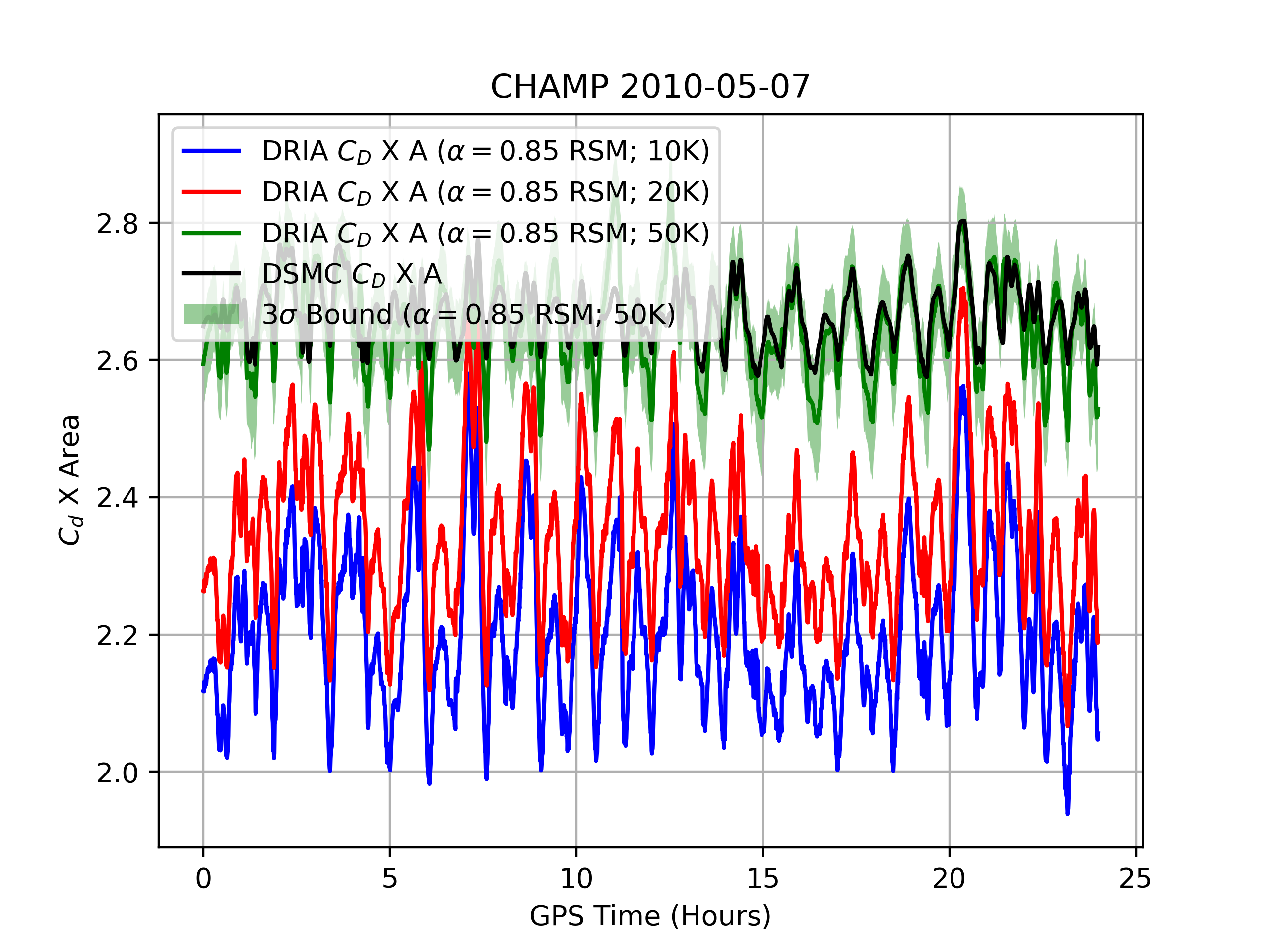} }}
    \subfloat[24-hour prediction for July 07, 2010]{{\includegraphics[width=0.45\textwidth]{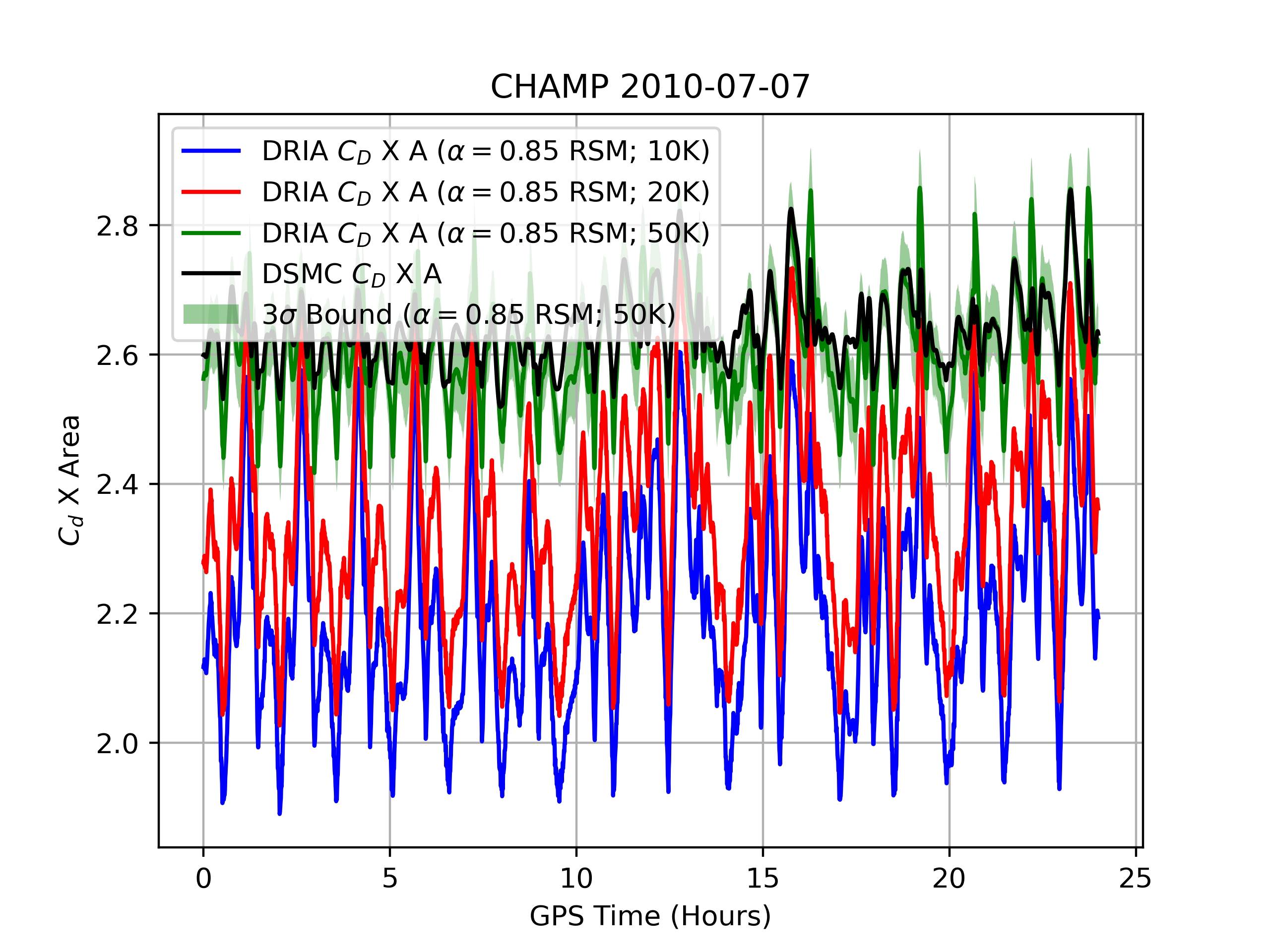} }} 
    \caption{CHAMP $C_D\times Area$ for four select dates for models trained using 10,000, 20,000, and 50,000 samples. Comparison of neural network predicted drag coefficient with SPARTA-based drag coefficient. }
    \label{fig:comparison_us_dsmc}
\end{figure*}

Figure \ref{fig:comparison_us_dsmc} shows a 24-hour comparison between the neural network predicted drag coefficient and the SPARTA-based drag coefficient for four randomly selected days - (a) March 27, 2002, (b) April 20, 2010, (c) May 07, 2010, (d) July 07, 2010 - for the CHAMP satellite. In each of the figures, we also show how the performance varies for models trained using different data sizes. The solid lines show the neural network predicted mean drag coefficient multiplied by the projected area of the CHAMP satellite. Blue color: the neural networks are trained using 10,000 samples; red color: the neural networks are trained using 20,000 samples; green color: the neural networks are trained using 50,000 samples. In black, we have the DSMC-based drag coefficient multiplied by the projected area of the CHAMP satellite. The shaded green area shows the neural network predicted $3\sigma$ uncertainty multiplied by the projected area for models trained using 50,000 samples. Uncertainties are not shown for models trained with 10,000 or 20,000 samples for clarity/readability. As evident from Fig. \ref{fig:comparison_us_dsmc}, the appropriate number of training samples is 50,000, as using 50,0000 training samples results in much better accuracy (determined by closeness to the DSMC results) in comparison to 10,000 or 20,000 training samples.

\section{Scalable Gaussian Process Modeling For Large-Scale Drag Coefficient Modeling - Difficulties}
Traditional Gaussian Processes provide accurate mean drag coefficient predictions and highly calibrated uncertainty estimates. But they do not scale well with the number of samples or output dimension. Another drawback of the traditional GPR is that the trained models are often too large to be saved, making reusability a challenging task. Despite using powerful computers (GPU-enabled/multiple cores), a GPR model training with 50,000 training samples is unsuccessful due to memory constraints. GPR model training with 10,000/20,000 training samples is feasible, but the saved models are in gigabytes, rendering them useless for later use in orbit uncertainty propagation studies. Therefore, we explore scalable GPR techniques, which are approximate but faster versions of the full-scale Gaussian Processes. Based on the literature review, we explore the following three scalable Gaussian Processes - (1) Parametric Gaussian Processes (PGP) \citep{PGP2019}, (2) GPflow-based Sparse Variational Gaussian Process (SVGP) \citep{GPflow2020multioutput} and (3) local approximate Gaussian Process (laGP) regression \citep{sun_gramacy_2019}. For the PGP regression, the training is computationally expensive, lacks sufficient prediction accuracy, and reusability is a challenging task. Therefore, we do not give it further consideration. For the GPflow-based SVGP, reusability and accuracy are the issues. Therefore, we do not recommend using it. The primary challenge with laGP is that the trained approximate models cannot be saved for later use, as it is a transductive model. We seek an inductive model for reusability purposes. Based on our investigation, we conclude that scalable Gaussian Processes are not ideal for large-scale drag coefficient modeling, where a primary requirement is that the trained models be saved and re-loaded later for orbit uncertainty propagation investigations. An alternative, such as the neural network, is more suitable for our application.

\section{Effect of Physical Drag Coefficient Modeling Uncertainties on Orbital State Uncertainties}
In this section, first, we demonstrate the importance of physical drag coefficient modeling for orbit propagation. Following that, we study the effects of drag coefficient uncertainties on orbital state uncertainties.

\subsection{Demonstration of the Importance of Physical Drag Coefficient Modeling}\label{importance}
In orbit propagation, drag coefficients are often assumed to be constants for simplicity or lack of a computationally efficient method of calculating drag coefficients. To demonstrate the drawback of this assumption and the need for modeling the physical drag coefficient, we consider six test case studies, which are detailed in Table \ref{mot_cases}. In each of the six cases, we consider two scenarios: (i) in the first scenario, we propagate an object with a variable drag coefficient which we compute using the machine learning models, and (ii) in the second scenario, the same object is propagated with a constant drag coefficient obtained by averaging the drag coefficient values along the orbit from the first scenario. We then investigate the temporal evolution of the along-track difference between the two scenarios.

\begin{table*}
    \footnotesize
    \centering 
    \caption{Description of test cases to motivate the importance of physical drag coefficient modeling. The parameter `$D$' is the time in days since the initial epoch of orbit propagation. The arguments for the sine and cosine functions in the definition of pitch and yaw are in radians.}
   \label{mot_cases}
   \begin{tabular}{|l | l | l | l | l |} % Column formatting, 
      \hline
      Case No. & Object & Drag Coefficient & Attitude & C/S Area\\
      \hline 
      Case I-a      &  Sphere & Physical $C_{D_i}$ from machine learning & N/A & Constant\\
      Case I-b       & Sphere & Constant $C_D=\frac{1}{\text{(n}_{steps})}\sum_{orbit}C_{D_{i}}$s (case I-a) & N/A & Constant\\
      \hline
      Case II-a      &  CHAMP & Physical $C_{D_i}$ from machine learning & Pitch = yaw = $0\degree$ & Constant\\
      Case II-b       & CHAMP & Constant $C_D=\frac{1}{\text{(n}_{steps})}\sum_{orbit}C_{D_{i}}$s (case II-a) & Pitch = yaw = $0\degree$ & Constant\\
      \hline 
      Case III-a      &  CHAMP & Physical $C_{D_i}$ from machine learning & Pitch $=\sin{(100D)}\degree$, & fn.(attitude)\\
            &   &  & yaw $=5\cos{(100D)}\degree$ & \\
      Case III-b       & CHAMP & Constant $C_D=\frac{1}{\text{(n}_{steps})}\sum_{orbit}C_{D_{i}}$s (case III-a) & Pitch $=\sin{(100D)}\degree$, & fn.(attitude)\\
             &  &  & yaw $=5\cos{(100D)}\degree$ & \\      
      \hline       
      Case IV-a      &  CHAMP & Physical $C_{D_i}$ from machine learning & Pitch $=90\sin{(100D)}\degree$, & fn.(attitude)\\
            &   &  & yaw $=180(1+\cos{(100D)})\degree$ & \\
      Case IV-b       & CHAMP & Constant $C_D=\frac{1}{\text{(n}_{steps})}\sum_{orbit}C_{D_{i}}$s (case IV-a) & Pitch $=90\sin{(100D)}\degree$, & fn.(attitude)\\
             &  &  & yaw $=180(1+\cos{(100D)})\degree$ & \\      
      \hline 
      Case V-a      &  CHAMP & Physical $C_{D_i}$ from machine learning & Pitch $=\sin{(100D)}\degree$, & fn.(attitude)\\
            &   &  & yaw $=5\cos{(100D)}\degree$ & \\
      Case V-b       & CHAMP & Constant $C_D=\frac{1}{\text{(n}_{steps})}\sum_{orbit}C_{D_{i}}$s (case V-a) & N/A & Constant \\
      \hline     
      Case VI-a      &  CHAMP & Physical $C_{D_i}$ from machine learning & Pitch $=90\sin{(100D)}\degree$, & fn.(attitude)\\
            &   &  & yaw $=180(1+\cos{(100D)})\degree$ & \\
      Case VI-b       & CHAMP & Constant $C_D=\frac{1}{\text{(n}_{steps})}\sum_{orbit}C_{D_{i}}$s (case VI-a) & N/A & Constant\\
      \hline       
   \end{tabular}
\end{table*}
\normalsize

The test case objects are assumed to be in a high-inclination, near-zero eccentricity orbit with an altitude of around 400 km. The initial orbital elements for all the test cases are the same and are given in Table \ref{initial_state}. Relevant simulation parameters are given in Table \ref{sim_param}. It is important to note that the selected propagation period roughly corresponds to a geomagnetic storm during a solar maximum. For orbit integration, we use a modified version of Dormand and Prince's Runge-Kutta Method \citep{DP1980} (also referred to as the `RK45' integrator in Python's scipy.integrate package \citep{2020SciPy-NMeth}). The modified integrator uses a constant integration step size of 10 seconds rather than striving for specified absolute and relative tolerances. This modification was made because a variable step size integrator takes a long time to converge in the presence of a stochastic drag coefficient, whose value changes in every internal adjustment of a single call of the step size. The cross-sectional areas of the objects for test cases I and II are taken as 0.770981 $m^2$. For test cases III, IV, V-a, and VI-a, cross-sectional area (attitude-dependent) is obtained by applying SciPy's \textit{LinearNDInterpolator} \citep{2020SciPy-NMeth} to an area look-up table. The constant cross-sectional areas for test cases V-b and VI-b are obtained by averaging the cross-sectional area values along the orbit from test cases V-a and VI-a, respectively.

For all six test cases, for the machine learning-based drag coefficient computations, we use the predicted mean drag coefficients (and ignore the predicted standard deviations). For the drag coefficient computation, we use Eq. \ref{CD}, where the satellite surface temperature ($T_w$) is taken as 400 K, the atmospheric translation temperature ($T_{\infty}$) is obtained from the NRLMSISE-00 density model, the adsorbate energy accommodation coefficient ($\alpha_{ads}$) is taken as 1 \citep{MEHTA2022}, and the surface energy accommodation coefficient ($\alpha_{surf}$) is obtained as \citep{MEHTA2022}:

\begin{equation}\label{alphasurf}
    \alpha_{surf} = \frac{3\mu}{(1+\mu)^2};\;\;\;\;\mu = \frac{\sum_{k=1}^6 (\chi_k m_k)}{m_{surf}}
\end{equation}
where $\chi_k$ is the mole fraction of species $k$, $m_k$ is the mass of species $k$, and $m_{surf}$ is the mass of a particle that composes the surface lattice (=263.3223 amu \citep{MEHTA2022}). Mole fractions for the species $H$, $He$, $N$, $N_2$, $O$, $O_2$ are obtained from the NRLMSISE-00 density model.

\begin{table}
    \centering 
    \caption{Keplerian elements defining the initial position of the satellites.}
   \label{initial_state}
   \begin{tabular}{|l | l | l |} % Column formatting, 
      \hline
      Orbital Element & Values\\
      \hline 
      Semi-major axis      & 6778136.3000 m\\
      \hline
      Eccentricity       & 2.2150$\times 10^{-3}$\\
      \hline
      Inclination       & $87.2193\degree$ \\
      \hline
      True anomaly       & $274.4887\degree$ \\
      \hline
      Argument of perigee & $85.6397\degree$ \\
      \hline
      Right ascension of ascending node    &  $206.9785\degree$ \\
      \hline
   \end{tabular}
\end{table}

\begin{table}
    \centering 
    \caption{Simulation parameters for the test cases.}
   \label{sim_param}
   \begin{tabular}{|l | l | l |} % Column formatting, 
      \hline
      Parameter & Values/Description\\
      \hline 
      Initial epoch      &  00:00:00 UT, November 20, 2003 \\
      \hline
      Propagation period &  3 days \\
      \hline
      Perturbations       &  $J_2$, atmospheric drag  \\
      \hline
      Atmospheric       & NRLMSISE-00 \\
      density model & \citep{PHDA2002} \\
      \hline
      Source for $ap$, $Ap$,        & CelesTrak (\cite{F107source},  \\
      $F10.7$ solar radio flux & \cite{apsource})\\
      \hline
      Satellite mass & 500 kg \\
      \hline
   \end{tabular}
\end{table}

\begin{figure}
\centering
\includegraphics[width = .5\linewidth]{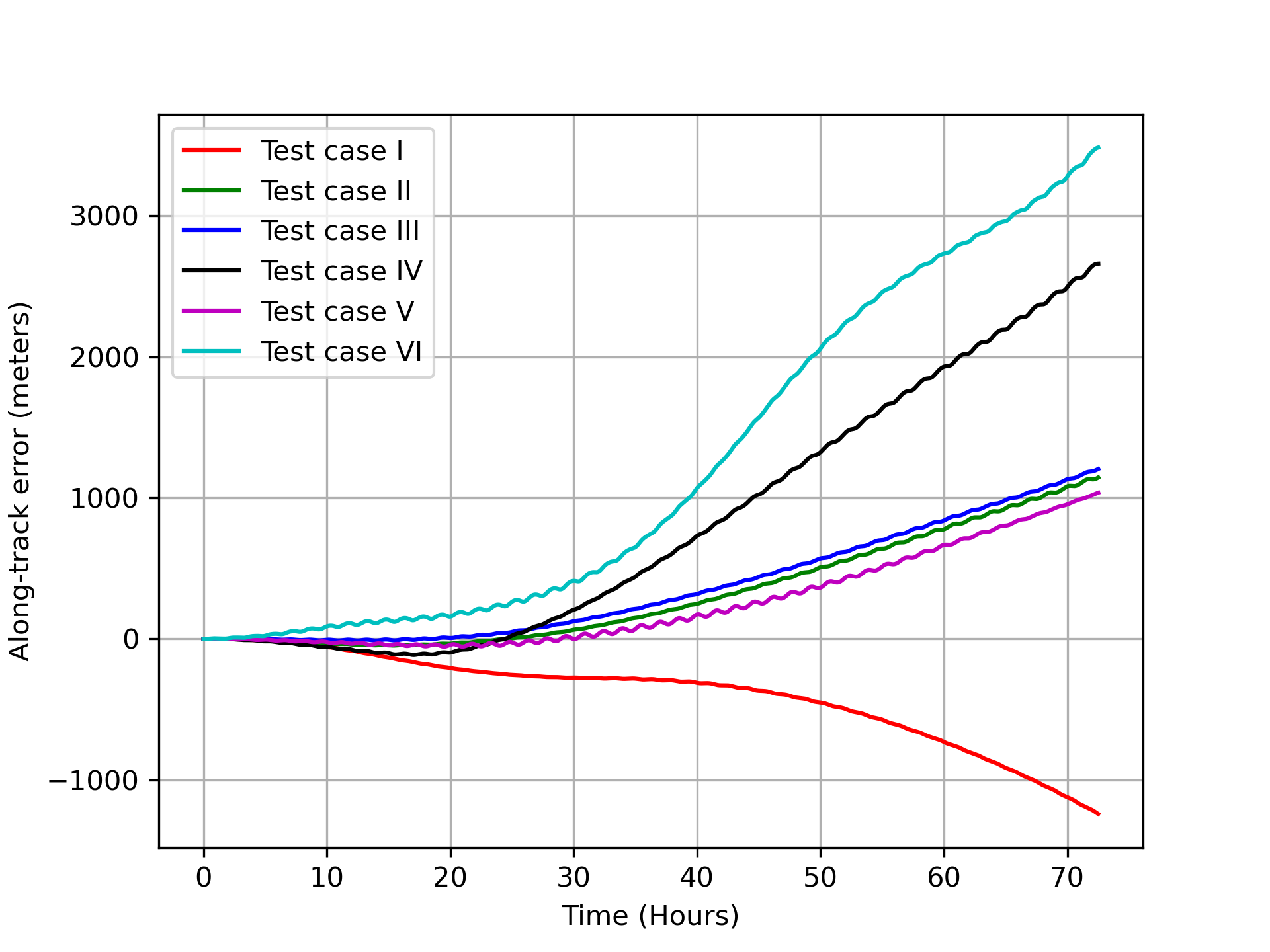}
\caption{Evolution of along-track error for different cases,  demonstrating the importance of physical drag coefficient modeling.}
\label{motivation}
\end{figure}

Figure \ref{motivation} shows the evolution of along-track errors for the six cases. Test cases I, II, III, IV, V, and VI are shown in red, green, blue, black, magenta, and cyan, respectively. The history of along-track errors shows that assuming a constant drag coefficient can result in errors of a kilometer or more at the end of the three-day propagation period.

\subsection{Orbital State Uncertainties}
We use the well-known Monte Carlo approach for orbit uncertainty propagation for the CHAMP satellite. Several different scenarios are investigated, where for each case, we use a total of 500 Monte Carlo runs and investigate the error distribution at the end of three days of orbit propagation. A top-level description of the Monte Carlo approach is given in Table \ref{toplevel_MC}. Uncertainty investigations are carried out for different (a) drag coefficient models, (b) spatiotemporal correlations for the drag coefficient, (c) attitude control profiles, (d) space weather conditions, and (e) altitudes.

\begin{table*}
\centering 
\caption{Top-Level algorithm for orbit uncertainty propagation.}
\label{toplevel_MC}
\begin{tabular}{l}
\hline
\textbf{Algorithm}: Monte Carlo simulations\\
\hline
For $j=1:500$\\
\hspace{10pt}$X_0=$ initial orbital state\\
\hspace{10pt}For $t = 0:10\;seconds:3 Days$\\ 
\hspace{20pt}For $s=1:6$ ($s$ represents the six species $H$, $He$, $N$, $N_2$, $O$, $O_2$)\\
\hspace{30pt} $\cdot$ $c_d$ ($j$,$t$,$s$) = a sampled $C_D$ value for species $s$ from its probability distribution function\\
\hspace{20pt}End loop\\
\hspace{20pt}$\cdot$ $C_D(j,t)=$ total drag coefficient computed using Eq. \ref{CD} and using the sampled drag coefficients of the six species\\
\hspace{20pt}$\cdot$ Using $C_D(j,t)$, compute the drag force. Propagate the object to the next time step\\
\hspace{10pt}End loop\\
\hspace{10pt}$X(j)$ = orbital state at the end of 3-day propagation\\
End loop\\
Compute the distribution of $X-X_{ref}$, where $X_{ref}$ is some reference orbit\\
\hline
\end{tabular}
\end{table*}

Unless otherwise stated, all the Monte Carlo simulations are run using the set-up described in Tables \ref{initial_state} and \ref{sim_param}. We use the same constant-step modified `RK-45' integrator described in section \ref{importance} for orbit propagation. Furthermore, drag coefficients are computed using machine learning models, and, similar to section \ref{importance}, the machine learning inputs $T_w$ = 400 K, $T_{\infty}$ is from the NRLMSISE-00 density model, $\alpha_{ads}$ = 1, and $\alpha_{surf}$ is given by Eq. \ref{alphasurf}.

\subsubsection{Different Drag Coefficient Models}\label{ddcm}
We consider three different drag coefficient models, the details of which are given in Table \ref{description_diffcdmodels}. Cases A and B are based on sampling from a normal distribution, whereas case C is based on a first-order Gauss-Markov process \citep{tapley}:

\begin{subequations}
\begin{equation}
    \kappa(t) = \exp{(-\eta \Delta t)} \kappa(t-\Delta t) + u_k(t) \sqrt{\frac{\Gamma^2}{2\eta}\big(1-\exp{(-2\eta \Delta t})\big)}
\end{equation}
\begin{equation}
    \eta = - \frac{\ln{0.5}}{\tau}
\end{equation}
\end{subequations}
where $u_k(t)$ is a random number sampled from the standard normal distribution. The factor $(\Gamma^2/(2\eta))$, which represents the steady-state variance of $\kappa$, is taken to be unity. The parameter $\tau$ is the ``half-life'' and governs the rate at which the auto-correlation fades. Based on literature \citep{McLaughlin2012}, the half-life $\tau$ is taken to be 1.8 minutes. At the initial epoch (i.e., $t=0$), $\kappa$ is simply a random number sampled from the standard normal distribution.

\begin{table*}
    \footnotesize
    \centering 
    \caption{Description of different drag coefficient models.}
   \label{description_diffcdmodels}
   \begin{tabular}{|l | l |} % Column formatting, 
      \hline
      Cases & $C_D$ Model (For Each Time Step Along The Orbit)\\
      \hline 
      A      &  Total $C_D$ sampled from $\mathcal{N}(C,.01)$, i.e., a normal distribution with constant mean $C$ and standard deviation .01.\\
      & \textit{Computation of $C$} -Let us consider case B described below. Let an object be propagated using $\mu_{C_D}$, which\\
      & varies along the orbit. Then the constant $C$ is the average of all the drag coefficient values along that orbit.\\
      \hline
      B       & For each species $i$, 500 drag coefficient values are sampled from $\mathcal{N}(\mu_i,\sigma_i)$. The mean $\mu_i$ and the standard\\
      & deviation $\sigma_i$ come from the machine learning models for species $i$. Total mean drag coefficient $\mu_{C_D}$ and the\\ 
      & associated standard deviation $\sigma_{C_D}$ are computed from $\mu_i$ ($i=1:6$) and the 500 samples using Eq. \ref{CD}.\\ & Total $C_D$ sampled from $\mathcal{N}(\mu_{C_D},\sigma_{C_D})$\\
      \hline
      C       &  Total mean drag coefficient $\mu_{C_D}$ and the associated standard deviation $\sigma_{C_D}$ are first computed in the same\\
      & manner as that of case B. A parameter $\kappa$ is then sampled from a first-order Gauss-Markov process. Total\\
      & drag coefficient is given as: $C_D=\mu_{C_D}+\kappa \sigma_{C_D}$\\
      \hline
   \end{tabular}
\end{table*}
\normalsize

Case C (Gauss-Markov process based) is the most realistic in operations as drag coefficients have spatiotemporal correlations. Case A (Gaussian noise-based), on the other hand, is the least realistic because the drag coefficient distribution changes over time depending on various input factors such as atmospheric temperature, the density of atmospheric species, and others. For all the three cases A, B, and C, we assume that the attitude varies as: pitch $=90\sin{(100D)}\degree$ and yaw $=180(1+\cos{(100D)})\degree$, where $D$ is the time in days since the initial epoch, and the argument for sine/cosine functions are in radians. The attitude-dependent cross-section area is obtained using \textit{LinearNDInterpolator}.

\begin{figure*}[!htbp]
    \centering
    \subfloat[Case A]{{\includegraphics[width=0.45\textwidth]{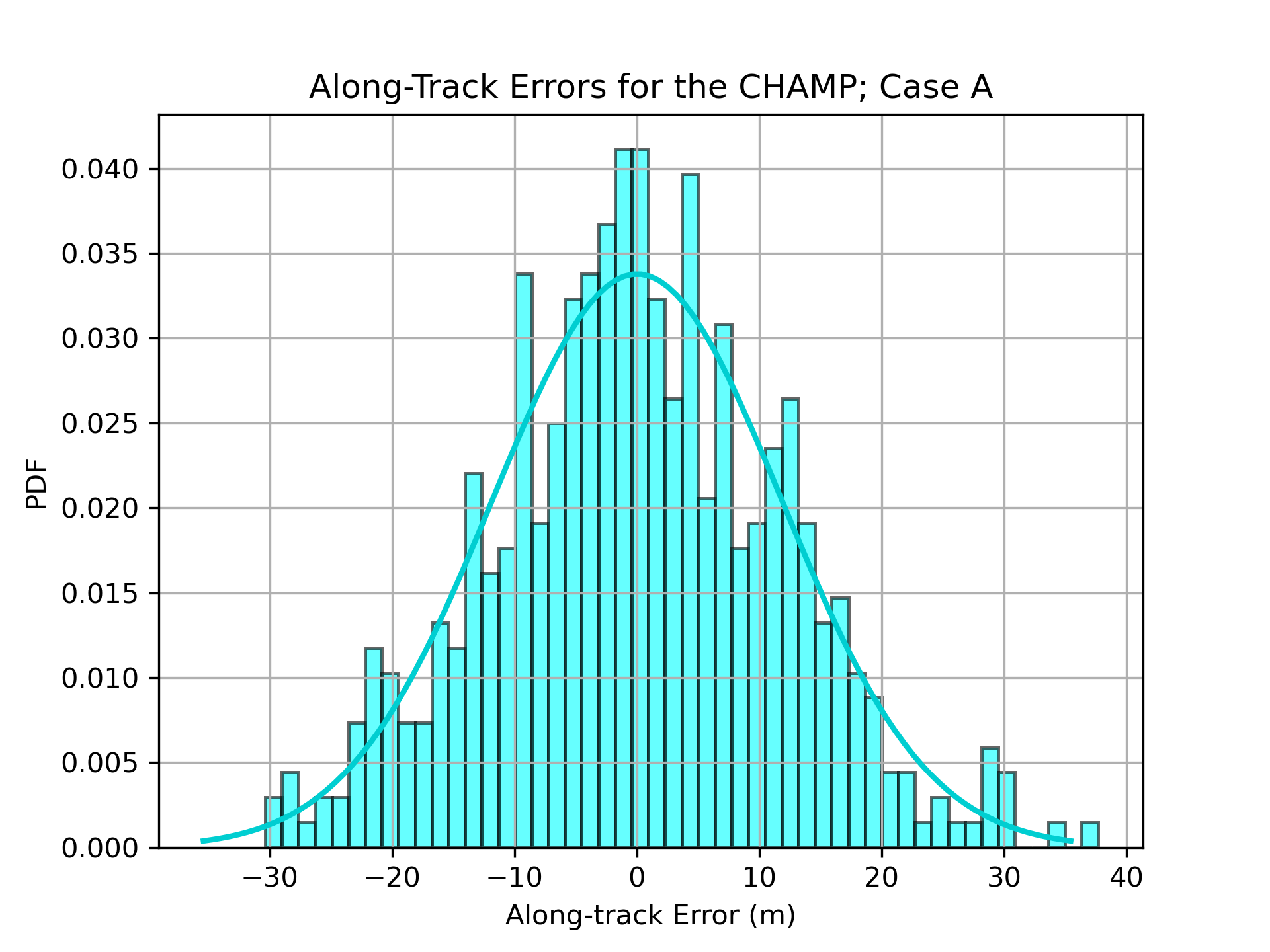}}}
    \subfloat[Case B and Case C]{{\includegraphics[width=0.45\textwidth]{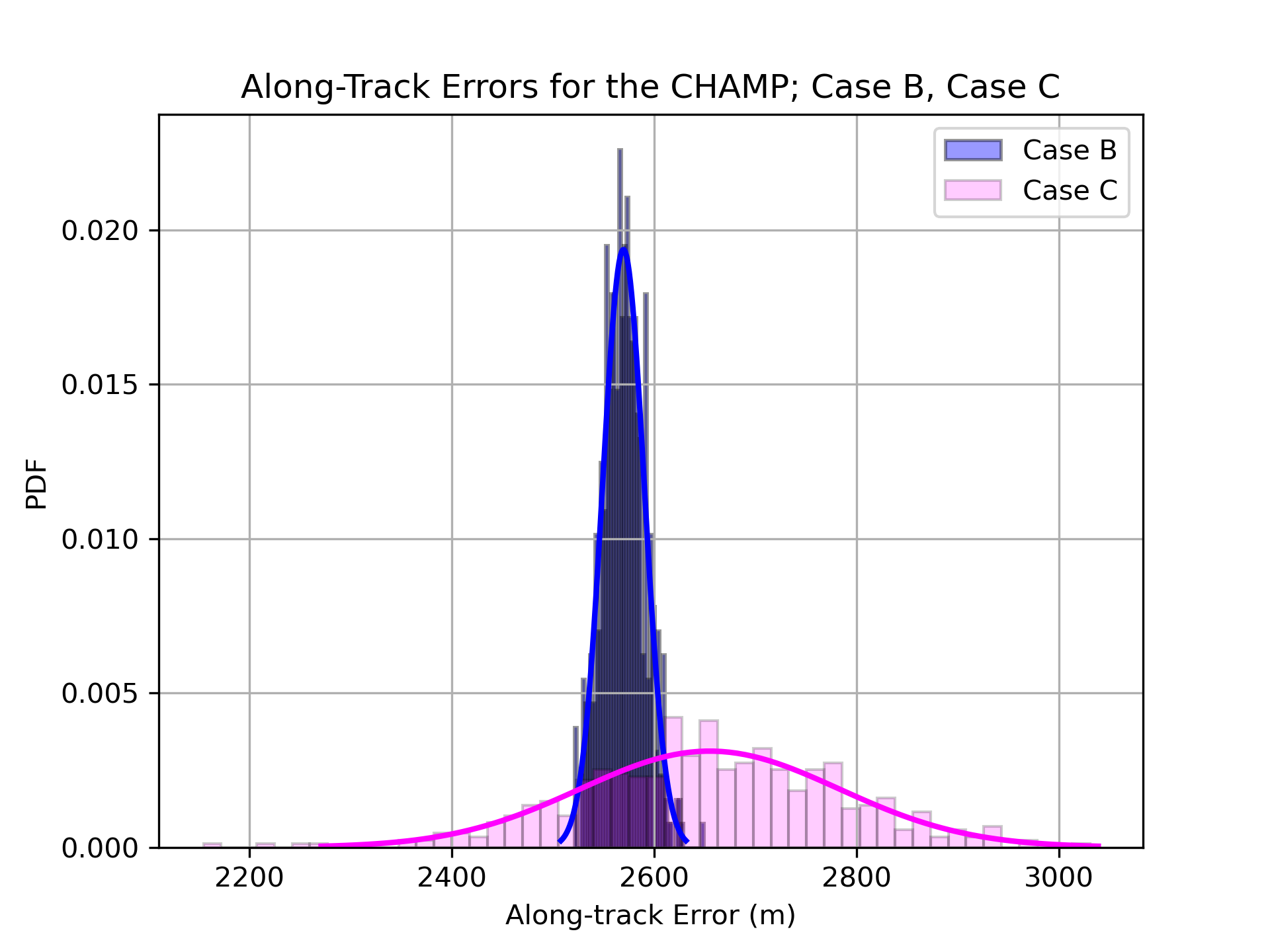} }}\\
    \caption{Along-track errors for cases A, B, and C at the end of three days of orbit propagation. Case A: Gaussian noise-based; Case B: machine learning-based normal distribution; Case C: based on Gauss-Markov process with half-life = 1.8 minutes. Attitude profile: full-attitude variation. Space weather condition: geomagnetic storm, solar maximum. Study of the effect of different drag coefficient models.}
    \label{along_track_diffcdmodels}
\end{figure*}

\begin{figure}
\centering
\includegraphics[width = .6\linewidth]{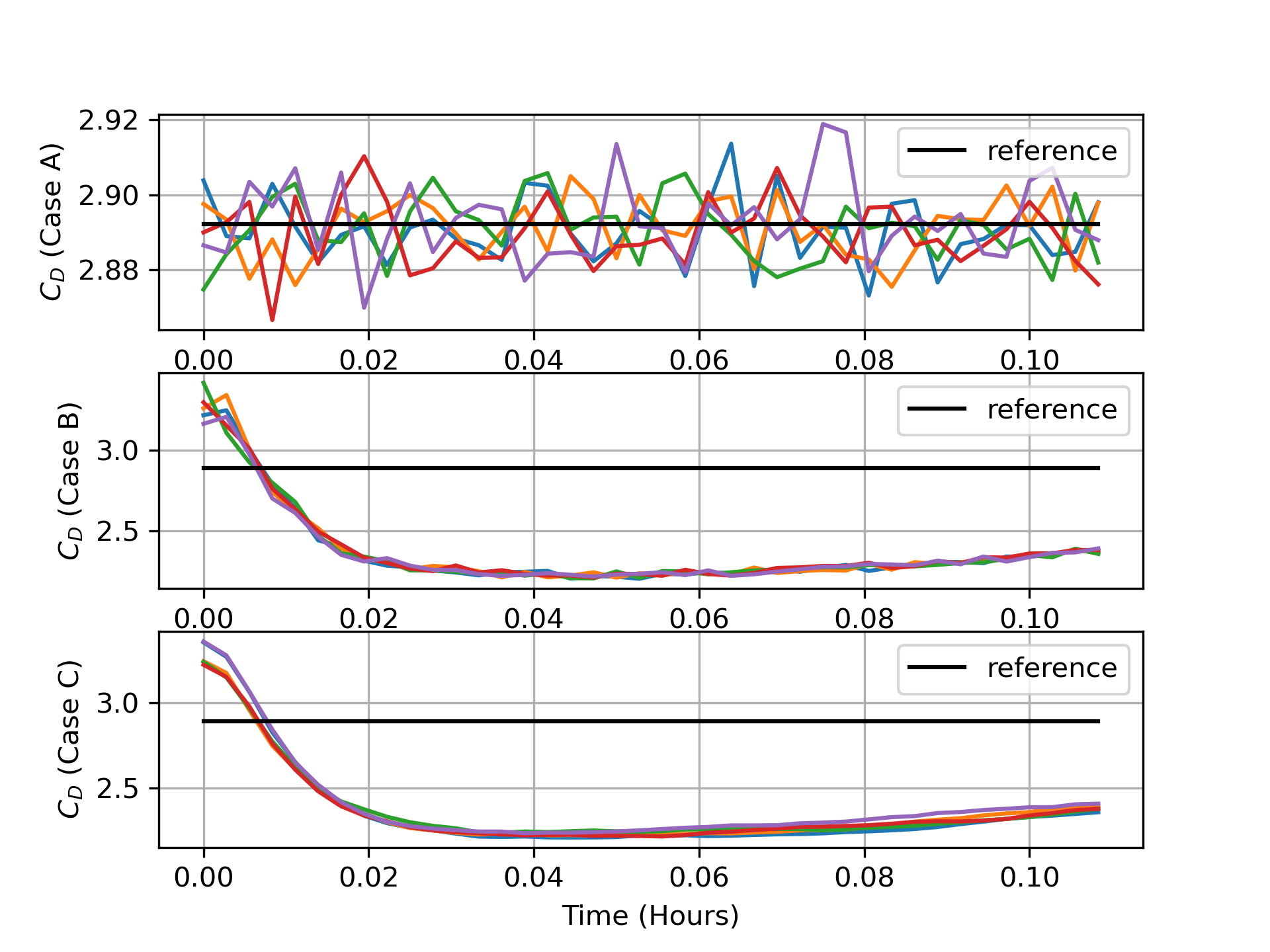}
\caption{Comparison of $C_D$ samples for different drag coefficient models. Case A: Gaussian noise-based; Case B: machine learning-based normal distribution; Case C: based on Gauss-Markov process with half-life = 1.8 minutes. Attitude profile: full-attitude variation. Space weather condition: geomagnetic storm, solar maximum. Note that for case B and case C, the mean drag coefficient values along the orbit (not shown in the figure) would lie close to the shown Monte Carlo samples; the variability of the $C_D$, i.e., the difference between the $C_D$ samples and the reference is not to be confused with the relatively small standard deviation values.}
\label{cdsamples_diffcdmodels}
\end{figure}

Figure \ref{along_track_diffcdmodels} shows the along-track errors at the end of three days of orbit propagation for cases A, B, and C. The reference orbit for the computation of the along-track errors is the orbit propagated with constant drag coefficient $C$ (see Table \ref{description_diffcdmodels} for details of $C$) with cross-sectional area varying according to the full-attitude profile described earlier. In Fig. \ref{along_track_diffcdmodels}, we show both the normalized histogram and the theoretical normal probability density function (PDF) fit. Compared to case A (Gaussian noise-based), the spread in the PDF (or the standard deviation) is much larger for case C (Gauss-Markov process-based). In case A, there is little to no spatiotemporal correlation in the sampled drag coefficient values along the orbit (partial random behavior); this results in the cancellation of perturbation effects, resulting in unrealistically small orbital errors. This behavior is demonstrated in Fig. \ref{cdsamples_diffcdmodels}, where we show the drag coefficients for the first five Monte Carlo samples and the reference orbit for the initial 390 seconds. Table \ref{bias_std_diffcdmodel} lists the bias and the $3\sigma$ uncertainties for the radial, along-track, and cross-track errors for all three cases at the end of three days of orbit propagation. From the table, the general trend is: $3\sigma$ for case C (Gauss-Markov process-based) $\gg$ $3\sigma$ for case B (machine learning-based normal distribution) $>$ $3\sigma$ for case A (Gaussian noise-based). The bias is approximately zero for case A and is of a similar order for cases B and C. Note that we do not show the plots for radial and cross-track errors because they are much smaller than the along-track errors and for brevity.

\begin{table*}[!htbp]
    \centering 
    \caption{Bias and $3\sigma$ uncertainties for the radial, along-track, and cross-track errors for cases A, B, and C at the end of orbit propagation. Case A: Gaussian noise-based; Case B: machine learning-based normal distribution; Case C: based on Gauss-Markov process with half-life = 1.8 minutes. Attitude profile: full-attitude variation. Space weather condition: geomagnetic storm, solar maximum.}    
   \label{bias_std_diffcdmodel}
   \begin{tabular}{|l | l | l |} % Column formatting, 
      \hline
      Errors & Bias (Case A $|$ Case B $|$ Case C) & Uncertainty ($3\sigma$ Values) (Case A $|$ Case B $|$ Case C)\\
      \hline 
      Radial (m) &  0.0039 $|$ -4.1968 $|$ -4.5222 & 0.1794 $|$ 0.2797 $|$ 1.8238 \\
      \hline
      Along-track (m) &  0.0136 $|$ 2569.3938 $|$ 2654.6444 & 35.4259 $|$ 61.7800 $|$ 384.0162 \\
      \hline
      Cross-track (m) &  1.1563E-5 $|$ 0.2362 $|$ 0.1302 & 0.0034 $|$ 0.0062 $|$ 0.1772 \\
      \hline
   \end{tabular}
\end{table*}

\subsubsection{Different Spatiotemporal Correlations for the Drag Coefficient}
\begin{figure}
\centering
\includegraphics[width = .5\linewidth]{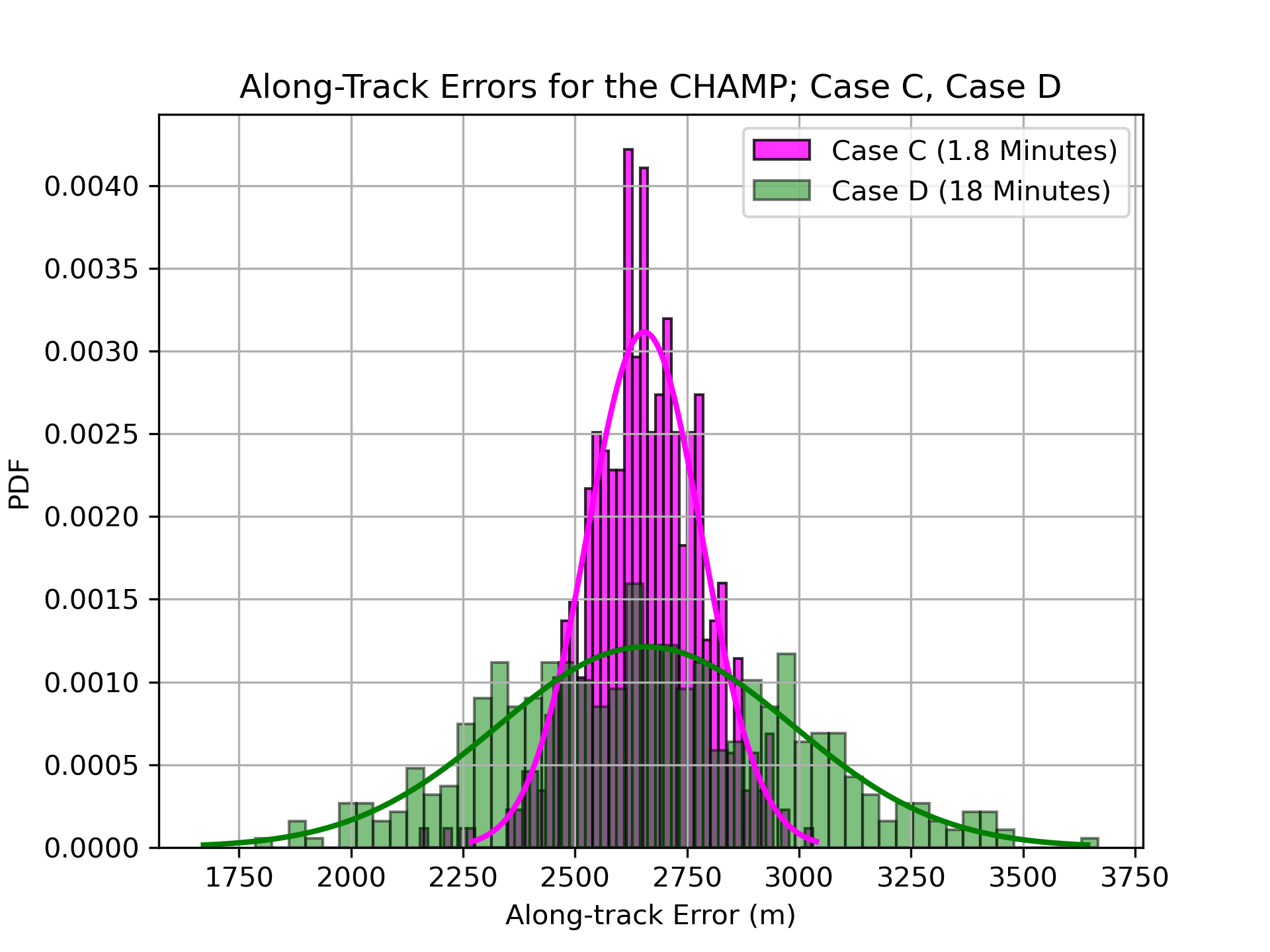}
\caption{Along-track errors for cases C and D at the end of three days of orbit propagation. Study of the effect of different drag coefficient spatiotemporal correlation. Case C/D: Gauss-Markov process-based. Attitude profile: full-attitude variation. Space weather condition: geomagnetic storm, solar maximum.}
\label{compare_casec_cased}
\end{figure}

In the first-order Gauss-Markov process-based orbit uncertainty propagation, i.e., case C (see Table \ref{description_diffcdmodels}), the half-life is taken as 1.8 minutes. However, in operations, the half-life can vary. We, therefore, re-run case C with a larger half-life of 18 minutes; we refer to this new case as `case D.' Figure \ref{compare_casec_cased} shows the distribution of along-track errors for cases C and D. The reference orbit we use here is the same as the one we use in section \ref{ddcm}. The spread (uncertainty) is much larger for case D (half-life = 18 minutes) because the sampled drag coefficient values are more correlated along the orbit, i.e., there is less random behavior in comparison to case C (half-life = 1.8 minutes). Tables \ref{bias_std_diffcdmodel} (for case C) and \ref{bias_std_cased} (for case D) list the bias and $3\sigma$ uncertainties for the radial, along-track, and cross-track errors at the end of orbit propagation. The bias is nearly identical for cases C and D.

\begin{table}[!htbp]
    \centering 
    \caption{Bias and $3\sigma$ uncertainties for the radial, along-track, and cross-track errors for case D at the end of orbit propagation. Case D: based on Gauss-Markov process with half-life = 18 minutes. Attitude profile: full-attitude variation. Space weather condition: geomagnetic storm, solar maximum.}    
   \label{bias_std_cased}
   \begin{tabular}{|l | l | l |} % Column formatting, 
      \hline
      Errors & Bias & Uncertainty ($3\sigma$ Values))\\
      \hline 
      Radial (m) &  -4.5428 &  4.0219\\
      \hline
      Along-track (m) &  2657.4929 & 986.7020 \\
      \hline
      Cross-track (m) &  0.1299 &  0.2157\\
      \hline
   \end{tabular}
\end{table}

\subsubsection{Different Attitude Control Profiles}
\begin{figure*}[!htbp]
    \centering
    \subfloat[Case E]{{\includegraphics[width=0.45\textwidth]{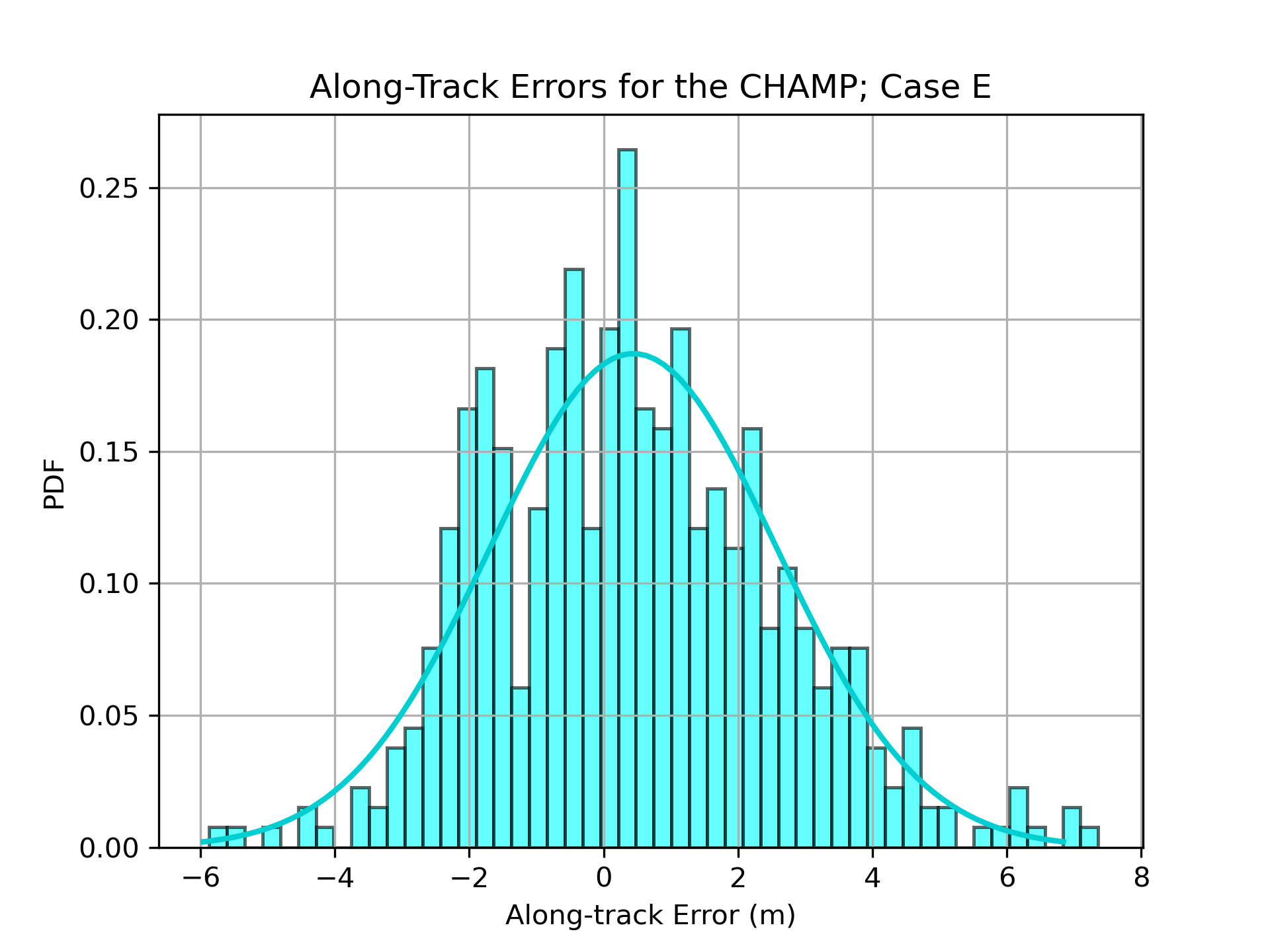}}}
    \subfloat[Case F and Case G]{{\includegraphics[width=0.45\textwidth]{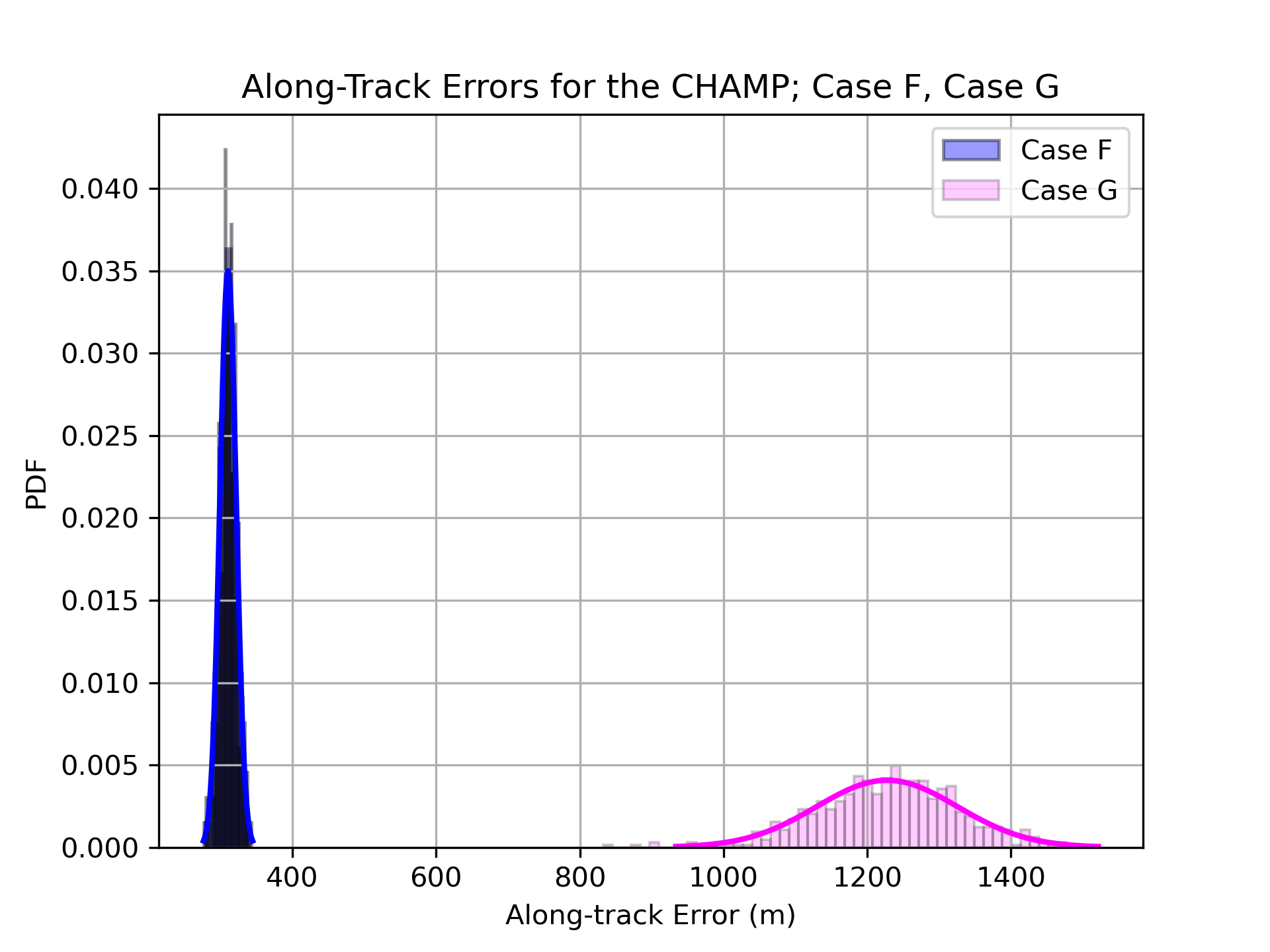} }}\\
    \caption{Along-track errors for cases E, F, and G at the end of three days of orbit propagation. Case E: Gaussian noise-based; Case F: machine learning-based normal distribution; Case G: based on Gauss-Markov process with half-life = 1.8 minutes. Attitude profile: reduced-attitude variation. Space weather condition: geomagnetic storm, solar maximum. Study of the effect of different attitude control profile.}
    \label{along_track_diffatt}
\end{figure*}

\begin{figure*}[!hbp]
    \centering
    \subfloat[Case C]{{\includegraphics[width=0.45\textwidth]{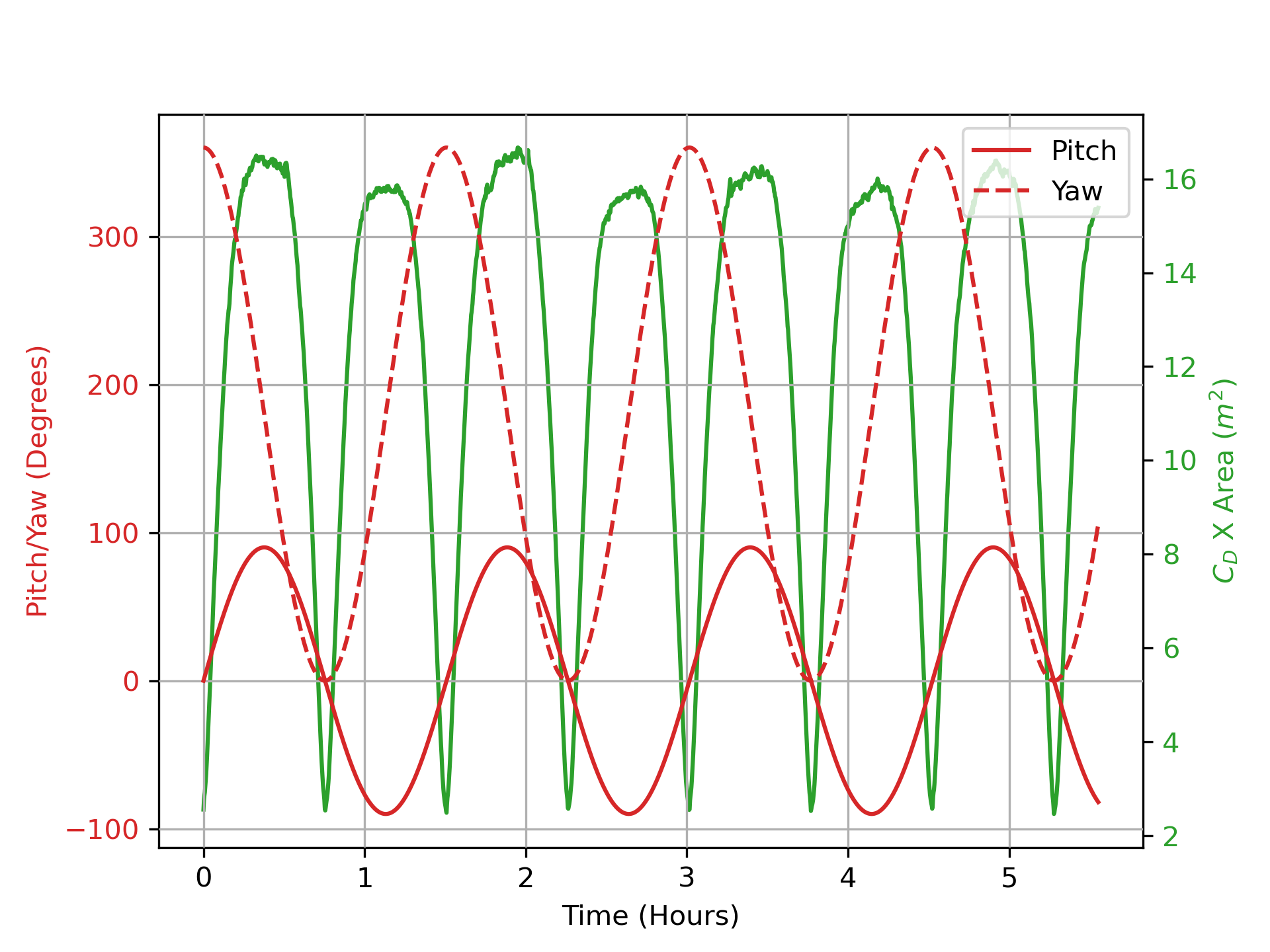}}}
    \subfloat[Case G]{{\includegraphics[width=0.45\textwidth]{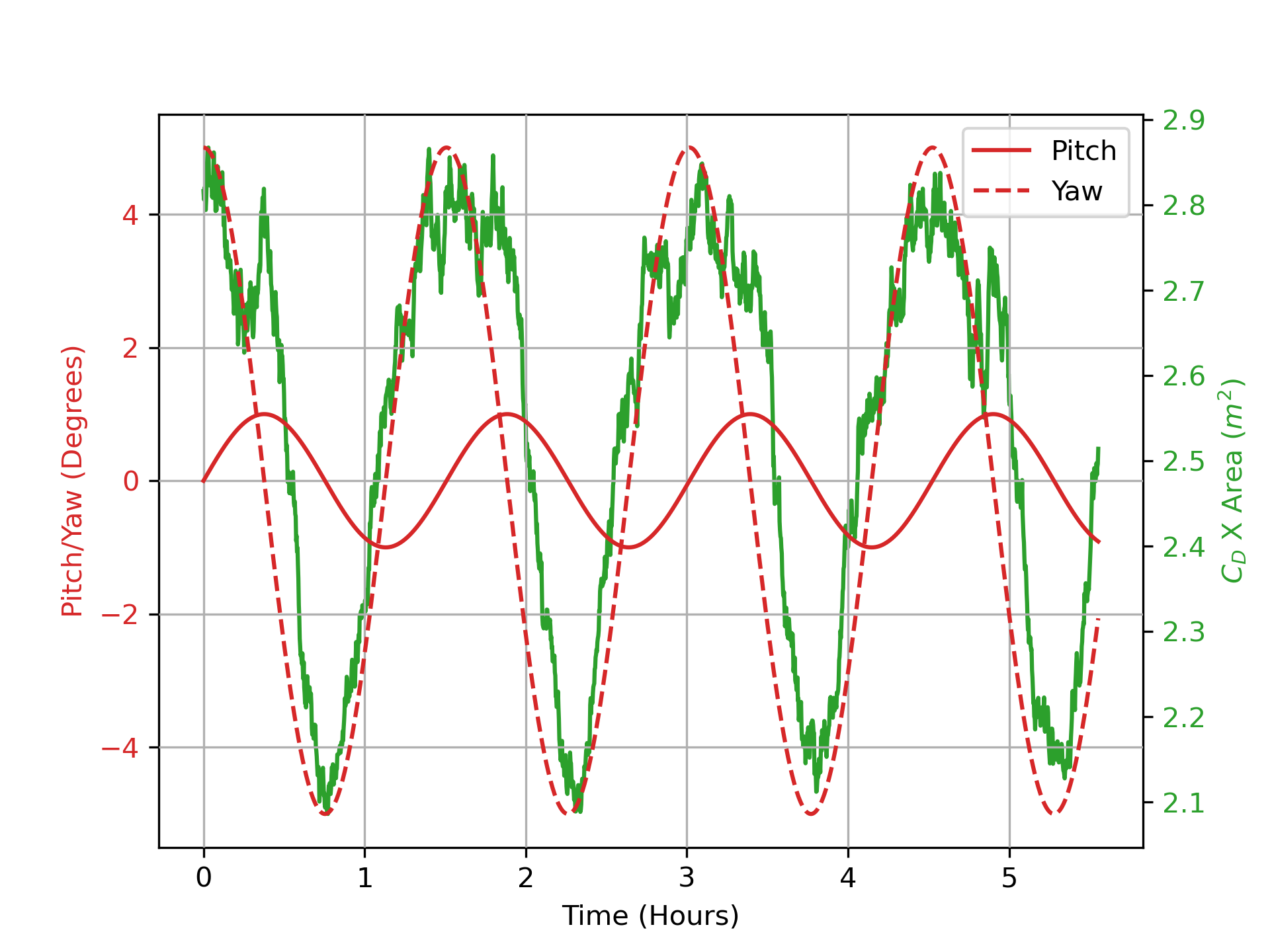} }}\\
    \caption{Comparison of ballistic coefficient for two randomly selected Monte Carlo samples from cases C and G, respectively. Case C: based on Gauss-Markov process with half-life = 1.8 minutes and full-attitude variation; Case G: based on Gauss-Markov process with half-life = 1.8 minutes and reduced-attitude variation. Space weather condition: geomagnetic storm, solar maximum.}
    \label{cd_times_area_diffatt}
\end{figure*}

For cases A, B, and C in section \ref{ddcm}, we assume a full-attitude variation (one can think of it as full-attitude control). However, space objects may have reduced-attitude control through devices such as reaction wheels, control moment gyros, and others. We, therefore, re-run cases A, B, and C with the following attitude profile: pitch $=\sin{(100D)}\degree$, yaw $=5\cos{(100D)}\degree$; we refer to these new cases as `case E', `case F', and `case G', respectively. Figure \ref{along_track_diffatt} shows the along-track errors for cases E, F, and G at the end of orbit propagation. For Fig. \ref{along_track_diffatt}, we generate the reference orbit using the same procedure as in section \ref{ddcm}, except with a reduced-attitude profile instead of the full-attitude profile. Table \ref{bias_std_diffatt} lists the bias and 3$\sigma$ uncertainties for the radial, along-track, and cross-track errors at the end of orbit propagation for cases E, F, and G. The $3\sigma$ values are smaller for the reduced-attitude cases when compared to the full-attitude cases. This is because of the smaller variations in ballistic coefficient along the orbit for the reduced-attitude cases; this is demonstrated in Fig. \ref{cd_times_area_diffatt}, where we compare $C_D \times Area$ for two randomly selected Monte Carlo samples - one from the full-attitude case (Case C) and one from the reduced-attitude case (case G) - for approximately six hours.

\begin{table*}[!htbp]
    \centering 
    \caption{Bias and $3\sigma$ uncertainties for the radial, along-track, and cross-track errors for cases E, F, and G at the end of orbit propagation. Case E: Gaussian noise-based; Case F: machine learning-based normal distribution; Case G: based on Gauss-Markov process with half-life = 1.8 minutes. Attitude profile: reduced-attitude variation. Space weather condition: geomagnetic storm, solar maximum.}    
   \label{bias_std_diffatt}
   \begin{tabular}{|l | l | l |} % Column formatting, 
      \hline
      Errors & Bias (Case E $|$ Case F $|$ Case G) & Uncertainty ($3\sigma$ Values) (Case E $|$ Case F $|$ Case G)\\
      \hline 
      Radial (m) & -0.0013  $|$ -2.1898 $|$ -7.7465 & 0.0329 $|$ 0.1642 $|$ 1.5581 \\
      \hline
      Along-track (m) & 0.4423 $|$ 310.4729 $|$ 1227.5856 & 6.3985 $|$ 34.1737 $|$ 293.9133\\
      \hline
      Cross-track (m) & 3.4203E-5 $|$ 0.0365 $|$ 0.0415 & 0.0007 $|$ 0.0035 $|$ 0.1878\\
      \hline
   \end{tabular}
\end{table*}

\subsubsection{Different Space Weather Conditions}
\begin{figure}[!hbp]
\centering
\includegraphics[width = .5\linewidth]{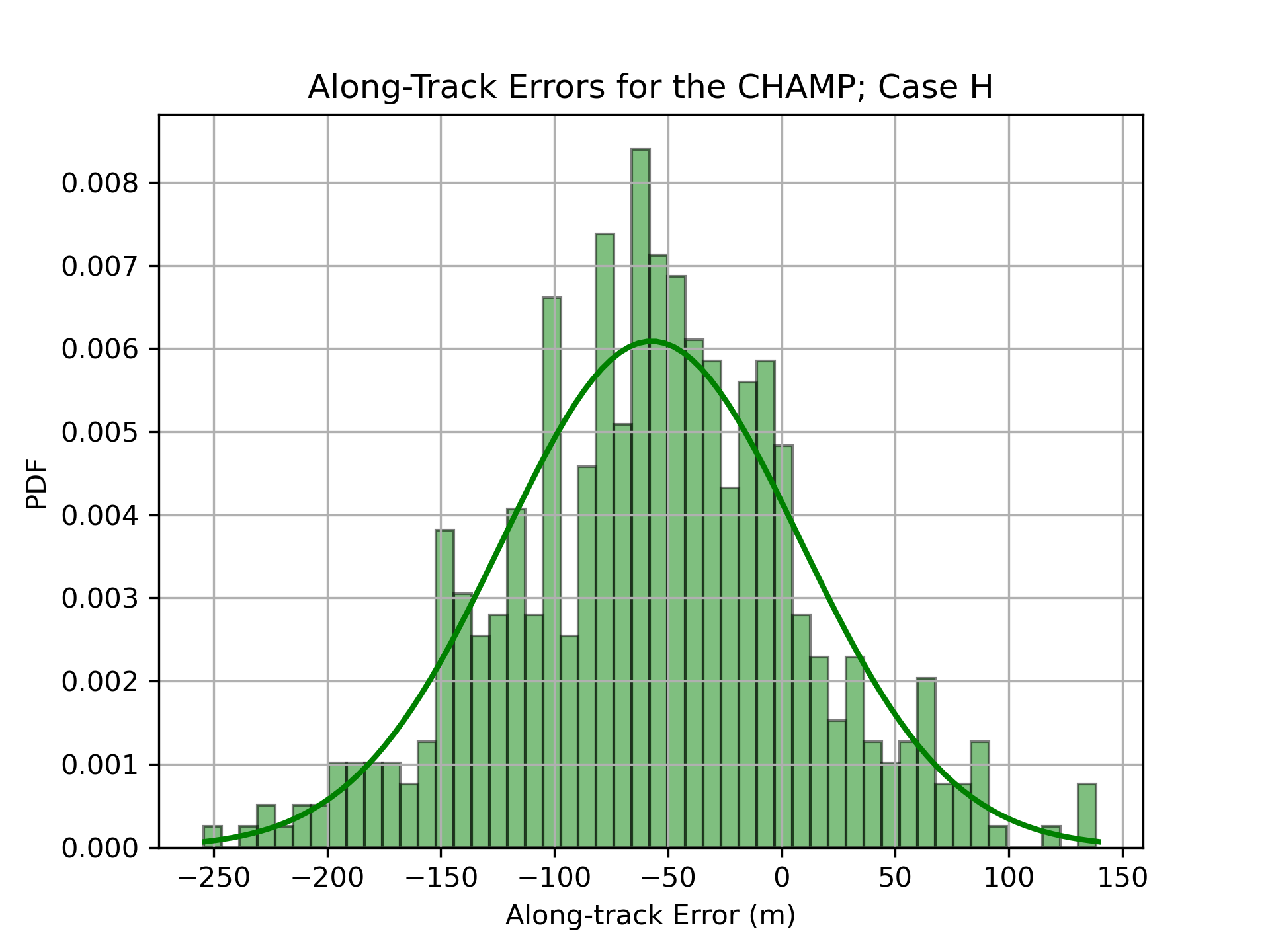}
\caption{Along-track errors for case H at the end of three days of orbit propagation. Case H: based on Gauss-Markov process with half-life = 1.8 minutes. Attitude profile: full-attitude variation. Space weather condition: quiet, solar minimum.}
\label{along_caseH}
\end{figure}

The effect of the drag coefficient on orbital perturbations can be significantly influenced by space weather conditions \citep{PSMH2021}. In the previous simulations, the propagation period roughly coincides with a geomagnetic storm during solar maximum. To study the effect of different space weather conditions, we re-run case C (Gauss-Markov process based) with an initial epoch of 00:00:00 UT, October 01, 2009, which corresponds to a quiet time during solar minimum; we refer to this new case as `case H.' Figure \ref{along_caseH} shows the along-track errors for case H at the end of orbit propagation. The reference for Fig. \ref{along_caseH} is an orbit generated by propagating an object with mean drag coefficients along the orbit computed from the machine learning models, full-attitude cross-sectional area variation, and quiet time space weather conditions. Table \ref{bias_std_caseH} lists the bias and $3\sigma$ uncertainties for the radial, along-track, and cross-track errors at the end of propagation for case H. The uncertainty in the along-track errors for the quiet time-based case H (196.5790 m) is almost half that of the uncertainty in the along-track errors for the storm time-based case C (384.0162 m). Compared to the geomagnetic storm/solar maximum period, the atmospheric densities are much smaller during the quiet solar minimum period. There is a strong coupling between the atmospheric density and drag coefficient values, which results in the differences between cases C and H. In any mission design, it is crucial to correctly model the space weather conditions to account for the correct impact of drag coefficient uncertainties.

\begin{table}
    \centering 
    \caption{Bias and $3\sigma$ uncertainties for the radial, along-track, and cross-track errors for case H at the end of orbit propagation. Case H: based on Gauss-Markov process with half-life = 1.8 minutes. Attitude profile: full-attitude variation. Space weather condition: quiet, solar minimum.}    
   \label{bias_std_caseH}
   \begin{tabular}{|l | l | l |} % Column formatting, 
      \hline
      Errors & Bias & Uncertainty ($3\sigma$ Values))\\
      \hline 
      Radial (m) & 0.3740  & 1.1161 \\
      \hline
      Along-track (m) & -57.2505 & 196.5790 \\
      \hline
      Cross-track (m) & 0.0501 & 0.1566 \\
      \hline
   \end{tabular}
\end{table}

\subsubsection{Different Altitudes}
\begin{figure}[!htp]
\centering
\includegraphics[width = .5\linewidth]{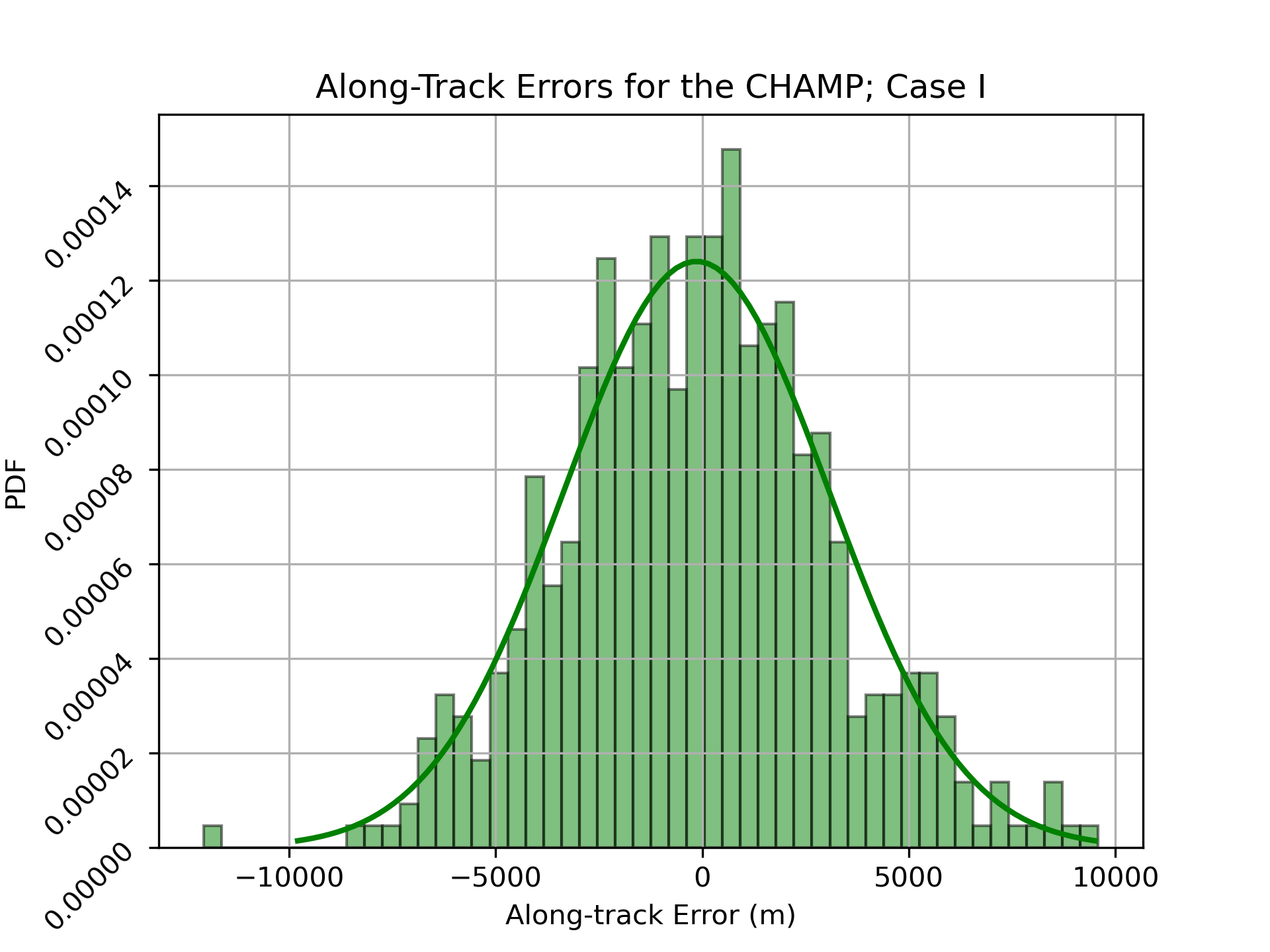}
\caption{Along-track errors for case I at the end of three days of orbit propagation. Case I: based on Gauss-Markov process with half-life = 1.8 minutes. Attitude profile: full-attitude variation. Altitude: approximately 250 km. Space weather condition: geomagnetic storm, solar maximum.}
\label{along_caseI}
\end{figure}

The simulations conducted so far assume an altitude of around 400 km. To investigate the impact of satellite altitude on drag coefficient-induced orbital uncertainties, we re-run case C with an initial semi-major axis of 6628136.3000 meters (altitude $\approx$ 250 km); we refer to this new case as `case I.' Since the starting position is different for case I (compared to all the previous simulation cases), a new reference orbit is defined with an initial location at an altitude of 250 km; we propagate the reference orbit with mean drag coefficients along the orbit computed from the machine learning models, full-attitude cross-sectional area variation, and storm time space weather conditions. Figure \ref{along_caseI} shows the along-track errors at the end of three days of orbit propagation for case I. Table \ref{bias_std_caseI} lists the bias and $3\sigma$ uncertainties for the radial, along-track, and cross-track errors at the end epoch for case I. Thus, at very low altitudes, the along-track uncertainties ($3\sigma$ values) can be of the order of 10 km, emphasizing the importance of taking drag coefficient uncertainties into account when conducting orbit propagation. Note that we did not investigate any further lower altitudes; at altitudes below 200 km, the free molecular flow assumption starts to break down, and our machine learning models are not valid in the continuum flow regime.

\begin{table}[!htp]
    \centering 
    \caption{Bias and $3\sigma$ uncertainties for the radial, along-track, and cross-track errors for case I at the end of orbit propagation. Case I: based on Gauss-Markov process with half-life = 1.8 minutes. Attitude profile: full-attitude variation. Altitude: approximately 250 km. Space weather condition: geomagnetic storm, solar maximum.}    
   \label{bias_std_caseI}
   \begin{tabular}{|l | l | l |} % Column formatting, 
      \hline
      Errors & Bias & Uncertainty ($3\sigma$ Values))\\
      \hline 
      Radial (m) & 1.3790  & 65.5767 \\
      \hline
      Along-track (m) & -140.2569 & 9651.4373 \\
      \hline
      Cross-track (m) & 0.1898 & 1.1521\\
      \hline
   \end{tabular}
\end{table}

\section{Conclusions and Future Possibilities}
In this paper, we use stochastic machine learning techniques to design surrogate models for predicting the physical drag coefficient for the complex CHAMP satellite. In our study, we demonstrate that the machine learning models, which are computationally much faster than the numerical methods, are able to emulate drag coefficient results with sufficient accuracy.

Unlike other surrogate models, such as the Gaussian Process regression, we particularly stress the computational efficiency, model accuracy, reliability of the predicted uncertainties, reusability of models, and applicability of the models to all attitudes in this study. We develop a methodology to determine the appropriate data size for training the feed-forward deep neural network models. With approximately 50,000 data points for training and another 50,000 data points for testing purposes, our models are able to predict drag coefficients for $H$, $He$, $N$, $N_2$, $O$, $O_2$ with root mean squared errors of 0.0295, 0.0188, 0.0250, 0.0159, 0.0142, 0.0209 and mean absolute calibration errors of 0.8845\%, 2.2404\%, 1.0279\%, 1.3170\%, 0.5117\%, and 0.8384\%, respectively. 

To establish the importance of drag coefficient uncertainties for space operations and space traffic management purposes, we carry out orbit uncertainty propagation via Monte Carlo simulations for a variety of (a) drag coefficient models, (b) spatiotemporal correlations for the drag coefficient, (c) attitude control profiles, (d) space weather conditions, and (e) altitudes. In orbit propagation, uncertainty in the drag coefficient mainly manifests itself in the form of uncertainty in along-track errors. We demonstrate a number of important observations - (1) uncertainty in along-track errors can be an order of magnitude higher for realistic sampling methods such as the first-order Gauss Markov process when compared to the naive sampling from a normal distribution, (2) the selection of correct values of parameters modeling the drag coefficient spatiotemporal correlation can significantly change the distribution of the orbital uncertainties, (3) orbital uncertainties are strongly affected by the attitude profile of the space object, (d) the effect of the drag coefficient uncertainties on orbital state uncertainties is much more significant under a geomagnetic storm during solar maximum when compared to a quiet period during solar minimum, and (e) the uncertainty in the along-track errors because of drag coefficient uncertainties can be of the order of 10 km (or more) for a low altitude of 250 km.

The current study only considers one gas-surface interaction model. In the future, we intend to explore different kinds of gas-surface interaction, which is the largest source of uncertainty in drag coefficient modeling. Additionally, note that the investigations conducted in this paper only concern themselves with the uncertainties emanating from fitting the stochastic models to the training data. In the future, we will also include the effects of the uncertainties in inputs, such as the atmospheric composition, temperature, and others resulting from the space weather uncertainties.

\section*{Acknowledgements}
The authors would like to acknowledge the use of the High Performance Computing Resources Thorny Flat at West Virginia University housed at the Pittsburgh Supercomputing Center, funded by the National Science Foundation (NSF) Major Research Instrumentation Program (MRI) Award \#1726534.

\bibliographystyle{unsrtnat}
\bibliography{refs_ascjpaper}  %%% Uncomment this line and comment out the ``thebibliography'' section below to use the external .bib file (using bibtex) .





% \end{thebibliography}

\end{document}